\documentclass[aps,pre,twocolumn,superscriptaddress,nofootinbib]{revtex4-2}
\usepackage{amssymb,graphicx,color,float}
\usepackage[intlimits]{amsmath}
\usepackage{amsmath}
\usepackage[english]{babel}
\usepackage[colorlinks]{hyperref}
\hypersetup{
colorlinks=blue,
linkcolor=blue,
filecolor=blue,
urlcolor=blue,
citecolor=blue
}
\pdfoutput=1
\graphicspath{{figures/}} 
\usepackage[normalem]{ulem}
\usepackage{comment}
\usepackage{ragged2e}
\usepackage{enumerate}
\usepackage[dvipsnames]{xcolor}
\usepackage{tikz}
\usepackage{soul}

\usetikzlibrary{shapes.geometric}
\usetikzlibrary{decorations.markings}
\usetikzlibrary{decorations.pathmorphing,patterns}
\usetikzlibrary{decorations}
\usetikzlibrary{calc}
\usetikzlibrary{intersections}
\usetikzlibrary{arrows}
\usetikzlibrary{matrix}
\usetikzlibrary{positioning}
\usetikzlibrary{external}
\graphicspath{{./}{./s/}}
\begin{document}

\title{Principal component analysis of absorbing state phase transitions }

\author{Cristiano Muzzi}
\affiliation{SISSA --- International School for Advances Studies, via Bonomea 265, 34136 Trieste, Italy}
\affiliation{INFN, Sezione di Trieste, via Valerio 2, 34127 Trieste, Italy}
\author{Ronald Santiago Cortes}
\affiliation{SISSA --- International School for Advances Studies, via Bonomea 265, 34136 Trieste, Italy}
\affiliation{ICTP --- The Abdus Salam International Centre for Theoretical Physics, Strada Costiera 11, 34151 Trieste, Italy}
\author{Devendra Singh Bhakuni}
\affiliation{ICTP --- The Abdus Salam International Centre for Theoretical Physics, Strada Costiera 11, 34151 Trieste, Italy}
\author{Asja~Jeli\'c}
\affiliation{ICTP --- The Abdus Salam International Centre for Theoretical Physics, Strada Costiera 11, 34151 Trieste, Italy}
\author{Andrea~Gambassi}
\affiliation{SISSA --- International School for Advances Studies, via Bonomea 265, 34136 Trieste, Italy}
\affiliation{INFN, Sezione di Trieste, via Valerio 2, 34127 Trieste, Italy}
\author{Marcello Dalmonte}
\affiliation{ICTP --- The Abdus Salam International Centre for Theoretical Physics, Strada Costiera 11, 34151 Trieste, Italy}

\author{Roberto Verdel}
\affiliation{ICTP --- The Abdus Salam International Centre for Theoretical Physics, Strada Costiera 11, 34151 Trieste, Italy}

\begin{abstract}
We perform a principal component analysis (PCA) of two one-dimensional lattice models belonging to distinct nonequilibrium universality classes -- directed bond percolation and branching and annihilating random walks with even number of offspring.
We find that the uncentered PCA of datasets storing various  system's configurations can be successfully used to determine the critical properties of these nonequilibrium phase transitions. In particular, in both cases, we 
obtain good estimates of the critical point and the dynamical critical exponent of the models. For directed bond percolation we are, furthermore, able to extract critical exponents associated with the correlation length and the order parameter. We discuss the relation of our analysis with low-rank approximations of datasets. 
\end{abstract}

\maketitle

\section{Introduction}

A significant milestone in theoretical physics has been the establishment of a classification of phase transitions in terms of universality classes at equilibrium, with field theory and renormalization group serving as major tools for understanding universality  \cite{wilson1983renormalization, fisher1998renormalization, cardy1996scaling, goldenfeld2018lectures}. While phase transitions at thermodynamic equilibrium are relatively well understood, a complete understanding of out-of-equilibrium phase transitions is much less developed, although the techniques of renormalization group have also been extended to nonequilibrium dynamics \cite{hohenberg1977theory, tauber2014critical}. 
Among the wide class of nonequilibrium systems, it is of particular interest to study those systems that are \textit{intrinsically} out of equilibrium, i.e., \ those that violate detailed balance. The main feature of these systems is the presence of probability currents between configurations \cite{zia2007probability}. Such systems include, for example, catalytic reactions and population dynamics and can be realized through nonequilibrium lattice models \cite{hinrichsen2000non, odor2004universality,marro2005nonequilibrium}. For these systems, no equilibrium counterpart is present. 

A recent and promising avenue to enhance our comprehension of phase transitions involves the use of data-driven
techniques, especially in their application to experimental contexts \cite{PhysRevResearch.5.013026, rem2019identifying}. Notably, in recent years, both supervised and unsupervised machine learning techniques have demonstrated their capability to identify and characterize equilibrium phase transitions, even in systems featuring intricate non-local order parameters \cite{mehta2019high, carleo2019machine,carrasquilla2017machine,PhysRevB.94.195105,PhysRevE.95.062122,PhysRevB.96.144432,PhysRevE.96.022140,PhysRevE.105.024121}.
Whether the efficacy of these techniques extends to systems that are out of equilibrium remains relatively unexplored, although some investigations in this direction have been recently pursued \cite{martiniani2019quantifying,seif2021machine, gillman2024combining, PatternPrl, PatternPrr, TangNatCom, PCPDML,PRLMLout,PRLNNNoneq,PRETransfer,SLDP, PhysRevB.109.075152,Turkeshi_2022}. 

The use of these methods not only presents intriguing challenges, but also raises profound questions. The essence of our understanding of phase transitions lies in fundamental concepts like coarse-graining and the renormalization group. These concepts have significantly influenced our grasp of universality, revealing that few, coarse-grained variables suffice to decipher long-distance behavior. An intriguing question arises: can this understanding be extrapolated to the realm of datasets, and approached through the lenses of data science? Namely, is the intricacy of out-of-equilibrium many-particle dynamics also falling within the picture of dimensional reduction - not only in its capacity of describing critical properties using simple order parameters, but also at the level of governing the full state evolution? If so, to which extent? We note that linearized dimensional reductions have already been shown to be informative about information content in other many-body contexts, such as  diagrammatic techniques \cite{zang2024machine}.

In this study, we address these questions and investigate the applicability and effectiveness of unsupervised learning methods to identify nonequilibrium phase transitions, with a specific focus on absorbing phase transitions \cite{henkel2008non}. We conduct a principal component analysis (PCA) \cite{jolliffe1990principal,Jolliffe} study of two paradigmatic lattice models belonging to distinct nonequilibrium universality classes: the directed bond percolation (DBP), in the directed percolation (DP) universality class, and the branching and annihilating random walks with even number of offspring (BARWe), in the parity conserving (PC) universality class. Our study demonstrates that PCA 
can be successfully used to get information on datasets describing nonequilibrium phase transitions. Specifically, we accurately identify the critical point and dynamical critical exponent for both DP and PC universality classes. Furthermore, in the case of DP, we successfully extract critical exponents related to the correlation length and the order parameter. 
A significant implication of these findings is the efficacy of low-rank approximations in 
identifying 
universality classes within datasets.

The rest of this work is organized as follows. In Sec.~\ref{sec:models}, we introduce the physical models considered in our study and the observables relevant to describe them. In 
Sec.~\ref{sec:methods}, we describe the techniques and the datasets used in our analysis. 
In  Sec.~\ref{sec:PCA_DP}
we present the results of our analysis for DBP, while in Sec.~\ref{sec:PCA_BARW}, we focus on the case of
BARWe. 
Finally, in Sec.~\ref{sec:conclusions}, we draw our conclusions and present 
possible outlooks.

\section{Models}
\label{sec:models}

Lattice models play an important role in understanding equilibrium phase transition. 
The equilibrium properties of these
models at inverse temperature $\beta$ are determined by an energy function, the Hamiltonian $H$, which characterizes the equilibrium distribution $P_s(\mathcal{C})\propto e^{-\beta H[\mathcal{C}]}$ of a configuration $\mathcal{C}$  
of the system. Non-equilibrium lattice models, instead, are defined by some dynamical evolution rules or, equivalently, by some set of transition probabilities $\{W(\mathcal{C}\to \mathcal{C}')\}$, which specify how the system evolves in time from configuration $\mathcal{C}$ to $\mathcal{C}'$ \cite{marro2005nonequilibrium}. Unlike
their equilibrium counterparts, the stationary probability distribution $P_s(\mathcal{C})$ of the configurations of these non-equilibrium lattice models is not known a priori. Indeed, $P_s(\mathcal{C})$ can only be
determined as the solution of a master equation, which in most of the cases of statistical systems cannot be solved exactly.  

In our study, we consider two lattice models defined in one spatial dimension which are driven out of equilibrium by the presence of an absorbing state, i.e., a state which cannot be left by the dynamics. The two models feature the same configuration space and have the same microscopic absorbing state,
but belong to different universality classes: directed bond percolation and branching and annihilating random walks with an even number of offspring. 
In both cases, the presence of an absorbing state implies that the system is out of equilibrium. Indeed, at equilibrium the transition rates from one configuration to another should satisfy the condition of
detailed balance $P_s(\mathcal{C})W(\mathcal{C} \to \mathcal{C}')=P_s(\mathcal{C}')W(\mathcal{C}' \to \mathcal{C})$ for every pair of configurations $\mathcal{C}$ and $\mathcal{C}'$. In the case of an absorbing state $\mathcal{C}_{\rm{abs}}$, the rate of leaving the absorbing configuration is $W(\mathcal{C}_{\rm{abs}}\to \mathcal{C}')=0$ for every $\mathcal{C}'$. For this reason, a system featuring an absorbing state cannot obey detailed balance\footnote{We point out that the existence of an absorbing state $\mathcal{C}_\textrm{abs}$ causes a violation of the detailed balance conditions only  
if the stationary distribution $P_s(\mathcal{C})$ is not concentrated on $\mathcal{C}_\textrm{abs}$, i.e., if  $P_s(\mathcal{C}) \neq \delta_{\mathcal{C},\mathcal{C}_{\textrm{abs}}}$. This concentration actually occurs
in systems of finite size and thus a genuine non-equilibrium state may emerge only in the thermodynamic limit. 
}, making it a genuinely out-of-equilibrium system. In the following, we specify the dynamics of the two models under consideration. 

\subsubsection{Directed Bond Percolation} 
Directed percolation is the simplest and most common example of universality class exhibited by systems undergoing a transition from an active to absorbing phase.
Critical behavior associated with DP universality manifests in various systems, including, but not limited to chemical reaction-diffusion models  \cite{PhysRevE.56.R6241}, the contact process \cite{harris1974contact}, damage spreading transitions \cite{grassberger1995damage}, surface growth with quenched disorder \cite{barabasi1995fractal}, Reggeon field theory \cite{janssen1981nonequilibrium, grassberger1981phase}, dissipative cold atoms \cite{marcuzzi2015non}, measurement-induced phase transitions  \cite{XhekAbsorbing,PiroliTriviality,MIPTClassical,IaconisLucasChen}. 

Being the most widespread universality class for nonequilibrium systems, we start our analysis with directed bond percolation, a nonequilibrium lattice model showing a transition in DP universality. We consider the evolution of DBP
on a one-dimensional lattice of length $L$. The state of the system at discrete time $t$ is specified by a configuration $\mathcal{C}_t=(x_{1,t},...,x_{L,t})$ where each $x_{i,t}$ takes values $0$ or $1$ and indicates, respectively, an active or inactive site. Let the parameter $p \in [0,1]$ be the percolation probability to the left and to the right. For a given configuration at time $t$, it is possible to determine the configuration at $t+1$ according to the following evolution rule 
\begin{equation}
    x_{i,t+1} = \begin{cases}
    1 & \text{if } x_{i-1,t}=1 \quad \text{and} \quad z^{-}_{i}<p, \\
    1 & \text{if } x_{i+1,t}=1 \quad \text{and} \quad z^{+}_{i}<p,  \\
    0 & \text{otherwise,}
\end{cases}
\end{equation}
where $z^{+/-}_i$ are i.i.d.~random numbers uniformly distributed in $[0,1]$. 

\begin{figure}[h]
\includegraphics[width=.45\textwidth]{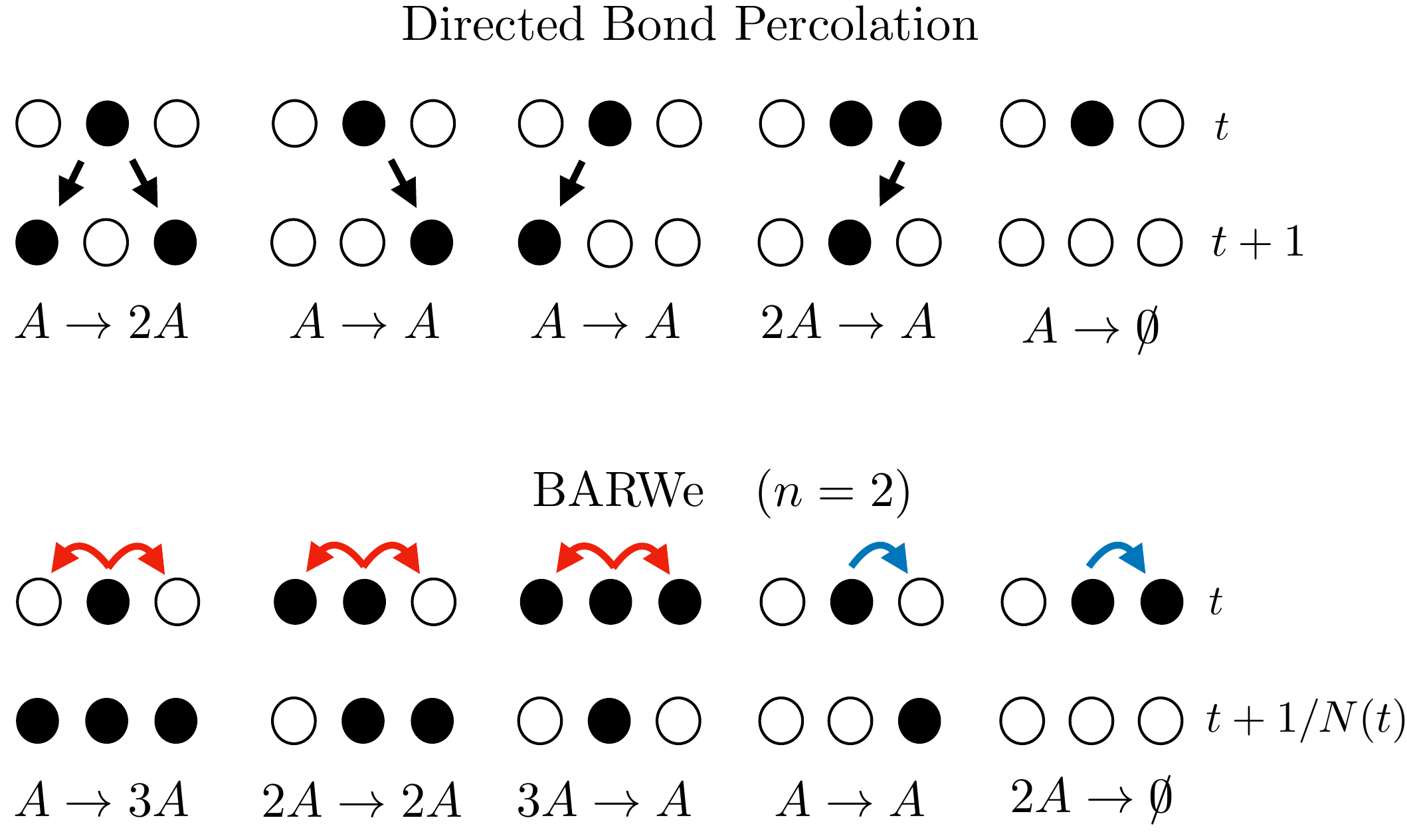}
\caption{Examples of moves for a $3$-sites lattice for DBP and BARWe $n=2$. For the BARWe, red arrows represent successful branching attempts, while blue arrows represent successful diffusion attempts. Notice that in the case of BARWe, the parity of the number of particles is conserved. 
}
\label{fig:moves}
\end{figure}

Such kind of dynamics can be interpreted as a reaction-diffusion process of a single particle species $A$  undergoing spontaneous decay $A \to \rlap{0}/$, branching $A \to 2 A$, and coagulation $2A \to A $, combined with single-particle diffusion. Examples of such processes are illustrated in Fig. \ref{fig:moves}. The ``vacuum'' state $\mathcal{C}_{\rm{vac}}=(0_1,...,0_L)$ is an absorbing state for the dynamics, since the spontaneous creation of particles (active sites) is not present. 

\subsubsection{Branching and Annihilating Random Walks}

Another well-established nonequilibrium universality class is the parity 
conserving one, which was first observed in Refs~\cite{grassberger1984new,Grassberger_1989}.
This universality class is the simplest possible deviation from the DP universality class and it emerges in various contexts,
including branching and annihilating random walks \cite{cardy1996theory}, nonequilibrium Ising models \cite{menyhard1994one},  systems with two absorbing states \cite{BasslerMonomer, HinrichsenSeveral}, roughening transitions \cite{RougheningPC}, and entanglement dynamics  \cite{odea2022entanglement, ravindranath2023entanglement}.

Among the models that belong to the PC universality class, we shall focus here on a special case of the branching and annihilating random walks (BARWs). The latter are  models of particles on a lattice which can diffuse, produce offspring on neighbouring sites, and annihilate upon contact. 
If interpreted as a reaction-diffusion process, this dynamics involves
a single species of diffusing particles which
may produce $n$ offsprings $A\to (n+1)A$ or annihilate in pairs $2A\to \rlap{0}/$ \cite{cardy1996theory}. For even $n$, this kind of dynamics conserves the number of particles modulo two, and thus given an initial state with an even (odd) particle number, the dynamics remains confined in the even (odd) parity sector. When the dynamics starts from a state with an even number of particles, the vacuum state is a reachable absorbing state for the dynamics.

Here we consider the microscopic model formulated in Ref.~\cite{zhong1995universality}, corresponding to $n=2$, i.e., a special case of the branching and annihilating random walk with even number of offsprings (BARWe). In this model, like in the previous one, the state of the system at time $t$ is specified by a configuration $\mathcal{C}_t=(x_{1,t},...,x_{L,t})$.  The update rule for the configuration is formulated as follows: A particle at site $i$ is randomly chosen. This particle makes an offspring production attempt on the two neighbouring 
sites with probability $p$ or a diffusion attempt to a randomly chosen neighbouring site with probability $1-p$. Whenever the neighbouring sites involved in the attempted offspring production or diffusion are occupied, the move is accepted with probability $k_a$, and annihilation of the neighbouring particles with the offspring occurs, 
while with probability $1-k_a$ the move is rejected. If instead, the attempted move involves only unoccupied neighbouring sites, the move is always accepted. 
Each time a particle is selected, a time counter is increased by $1/N(t)$, where $N(t)$ is the number of particles at time $t$, such that in a time step each particle either branches or diffuses a single time, on 
average. In our analysis we fix the annihilation rate to be  $k_a=\frac{1}{2}$.
%%
%%
%%%%%%%%%%%%%%%%%%%%%%%%%%%%%%%%%%%%%%%%%%%%%%%%%%%
%%%%%%%%%%%%%%%%%%%%%%%%%%%%%%%%%%%%%%%%%%%%%%%%%%%
%%%%%%%%%%%%%%%%%%%%%%%%%%%%%%%%%%%%%%%%%%%%%%%%%%%
%%%%%%%%%%%%%%%%%%%%%%%%%%%%%%%%%%%%%%%%%%%%%%%%%%%\
%%
%%

\subsubsection{Observables and Phase Transitions}

For both models, a pivotal quantity is the average density of active sites 
\begin{equation}\label{eq: def_density}
    \rho(t)=\biggr\langle \frac{1}{L}\sum^L_{i=1} x_{i,t} \biggr\rangle, 
\end{equation}
where $\langle \cdots \rangle$ denotes an average over an ensemble of realizations of the stochastic evolution.  
In the limit of infinite system size $L\to \infty$ and as $p$ is increased, both  models undergo
a continuous transition from the absorbing phase to the active phase. 
In the case of the BARW, this also requires having an even initial number of particles. 
This transition, occurring at a model-dependent value $p_c$ of $p$, can be detected by the stationary density of active sites 
\begin{equation}
    \rho_{\infty}=\lim_{t \to \infty} \rho (t)
\end{equation}
as an order parameter that is zero in the absorbing phase, while it attains a nonzero value in the active phase.
For values of $p$ close to the critical point $p_c$, the stationary density behaves as 
 \begin{equation}
     \rho_\infty \sim (p-p_c)^\beta \quad (p>p_c).
 \end{equation}
 When the dynamics starts from a fully active configuration, at the critical point $p=p_c$ the density of particles shows an algebraic decay
\begin{equation} \label{eq:decay_rho}
    \rho(t)\sim t^{-\alpha}.
\end{equation}
 In addition, it is possible to define correlation functions \cite{livi2017nonequilibrium}:

\begin{align}
    &C(|i-j|)= \lim_{t \to \infty} \big\langle \frac{1}{t} \sum^t_{\tau=0} (x_{i,\tau}-\overline{x}) (x_{j,\tau}-\overline{x}) \big\rangle,\\
    &C(t)=\lim_{L \to \infty} \big\langle  \frac{1}{L}\sum^L_{i=1} (x_{i,0}-\overline{x}) (x_{i,t}-\overline{x}) \big\rangle,
\end{align}
where we introduced $\overline{x}=\lim_{t \to \infty} \frac{1}{t} \sum^t_{\tau=0} \langle x_{i,\tau}\rangle$, which actually does not depend on the site index $i$. 
Far from criticality, these correlation functions decay exponentially as
\begin{align}
    &C(|i-j|) \sim e^{-|i-j|/\xi_{\bot}},\\
    &C(t)\sim e^{-t/\xi_{\parallel}},
\end{align}
and their exponential decay defines the correlation length $\xi_{\bot}$ and the correlation time $\xi_{\parallel}$.
Close to criticality $p\to p_c$, these quantities diverge as
 \begin{align}
     & \xi_{\bot}\sim |p-p_c|^{-\nu_{\bot}},\\
    &\xi_{\parallel}\sim |p-p_c|^{-\nu_{\parallel}}.
 \end{align}
 The anisotropy between the time and space directions is quantified by the dynamical critical exponent $z=\nu_{\parallel}/\nu_\bot$, which relates the divergence of the correlation time $\xi_\parallel$ to the one of the correlation length $\xi_\perp$ as $\xi_{\parallel}\sim \xi^z_\perp$. 

The transition from the absorbing to the active phase is known to belong to the DP universality class for DBP and to the PC universality class for the BARWe (investigated for $k_a=1/2$ in Ref.~\cite{zhong1995universality}.)
The critical points of both models in one spatial dimension $d=1$, together with the corresponding universal exponents of the DP and PC universality classes are reported in Table~\ref{table:tabexpDPPC}.
%%
%%
%%%%%%%%%%%%%%%%%%%%%%%%%%%%%%%%%%%%%%%%%%%%%
%%%%%%%%%%%%%%%%%%%%%%%%%%%%%%%%%%%%%%%%%%%%%
\begin{table}[h]
\footnotesize
\begin{center}
\begin{tabular}{|c|c|c|}
\hline\\[-4mm]
Quantity &  DBP $(d=1)$  \cite{hinrichsen2000non,JensenTemporallyDisordered} & BARW $n= 2$ ($d=1$) \cite{zhong1995universality} \\ \hline
$p_c$  & $ 0.6447001(1)$ &  $0.4946(2)  $\\
\cline{2-3}
 &  DP $(d=1)$ \cite{hinrichsen2000non,Jensen99a}  & PC $(d=1)$ \cite{jensen1994critical,Park_2013,grassberger2013continuum} \\ 
\cline{2-3}
$\beta$ & $0.276486(8)$ & $0.92(3)$   \\
$\nu_{\bot}$ &  $1.096854(4)$ & $1.84(6)$   \\
$\nu_{\parallel}$ &  $1.733847(6)$ &$3.25(10)$   \\
$z$ &  $1.580745(10)$ & $1.7415(5)$  \\ %\hline
$\alpha$ &  $0.159464(6)$ & $ 0.2872(1)$   \\
\hline
\end{tabular}
\end{center}
\caption{
Estimates of the values of the critical points $p_c$ for the directed bond percolation (DBP) and for the BARWe with $n=2$ \cite{zhong1995universality} and of the universal critical exponents for the corresponding universality classes, i.e., respectively,
directed percolation (DP) \cite{hinrichsen2000non,JensenTemporallyDisordered,Jensen99a} 
and parity conserving (PC) (the estimates for  $\beta,\nu_\bot,\nu_\parallel$ were obtained from the $n=4$ BARW in Ref.~\cite{jensen1994critical}, while the  estimates for $z$ and $\alpha$ are taken from \cite{Park_2013} for the $n=2$ BARW), in one spatial dimension. In $d=1$ both DP and PC have $\alpha=\delta$, where $\delta$ is the exponent characterizing the decay of the survival probability \cite{odor2004universality}.
}\label{table:tabexpDPPC}

\end{table}
%%%%%%%%%%%%%%%%%%%%%%%%%%%%%%%%%%%%%%%%%%%%%
%%%%%%%%%%%%%%%%%%%%%%%%%%%%%%%%%%%%%%%%%%%%%
%%
%%

In a numerical simulation, however, the system size $L$ is  finite. This implies that there is always a finite probability of ending in 
the absorbing state $\mathcal{C}_{\rm{vac}}=(0_1,...,0_L)$
and thus, in the infinite-time limit, the system always reaches it, i.e., $\rho_\infty=0$. Although a bona fide active stationary state exists only for $L\to \infty$, finite-size systems, in the active phase, display a quasistationary behavior: 
after a transient, physical quantities do not change significantly for some time, before attaining the values corresponding to the absorbing state. Upon increasing $L$, the duration of this quasistationary behavior rapidly increases and the observables converge to their stationary values. Close to criticality, the quasistationary state is reached from a generic initial condition in a timescale $t_c\sim L^z$. 
Accordingly, in order to study the active state from simulations of finite systems we focus on this quasistationary state by considering the observables at times $t\simeq t_c$.

\section{Methods and Datasets }\label{sec:methods}

\subsubsection{Principal Component Analysis}

Principal component analysis can be regarded as an application of the singular value decomposition (SVD) \cite{trefethen2022numerical} as a compression algorithm. Let us consider $N_r$ $m$-dimensional realizations (or observations) of a vector of $m$ variables $\vec{x}^{(\mu)}=(x^{(\mu)}_1,...,x^{(\mu)}_m) \in \mathbb{R}^m$, with $\mu=1,\ldots, N_r$. Each observation is a data point in $\mathbb{R}^m$.  In the case of a physical system, one can think of $m$ as the number of degrees of freedom which describe the state of the system. Given the ensemble of observations $\{\vec{x}^{(\mu)}\}^{N_r}_{\mu=1}$, 
we can introduce a $N_r\times m$ \textit{data matrix}
\begin{equation}
    \mathbf{X}=(\vec{x}^{(1)},...,\vec{x}^{(N_r)})^T,
    \label{eq:def-X}
\end{equation}
having the observations as rows. We introduce $\widetilde{\mathbf{X}}$, the \textit{centered data matrix}, as the matrix having components 
\begin{equation}
    \widetilde{\mathbf{X}}_{\mu i}=\mathbf{X}_{\mu i}-\frac{1}{N_r}\sum^{N_r}_{\mu'=1}\mathbf{X}_{\mu' i}.
    \label{eq:centering}
\end{equation}
$\widetilde{\mathbf{X}}$ is characterised by the fact that, contrary to $\mathbf{X}$, the arithmetic average of its rows vanishes. 
Starting from $\mathbf{X}$ and $\tilde{\mathbf{X}}$, one can construct two $m\times m$ matrices which encode information about correlations present in the dataset: the \textit{empirical covariance matrix}
\begin{equation}
    \mathbf{\Sigma}= \frac{1}{N_r} \widetilde{\mathbf{X}}^T
    \widetilde{\mathbf{X}},
    \label{eq:def-cov}
\end{equation}
and the \textit{empirical second-moment matrix}
\begin{equation}
    \mathbf{M}= \frac{1}{N_r} \mathbf{X}^T\mathbf{X}.
    \label{eq:def-M}
\end{equation}

For the sake of brevity, we will refer to $\mathbf{\Sigma}$ and $\mathbf{M}$ simply as covariance and second-moment matrices. 
 We denote the spectra of $\mathbf{\Sigma}$ and $\mathbf{M}$ as $\{\lambda^c_i\}$ and $\{\lambda^{u}_i\}$, respectively, with $i =1, \ldots, m$, where the eigenvalues are sorted in 
 non-increasing order, 
 i.e.,
 $\lambda^{c/u}_i\ge \lambda^{c/u}_{i+1}$.
 Let us notice that the number of nonzero eigenvalues of $\mathbf{\Sigma}$ or $\mathbf{M}$ is upper bounded by their rank. This rank is the same as the corresponding data matrix and thus it is smaller than or equal to 
 $\min(N_r,m)$. The procedure of computing eigenvalues and eigenvectors of either the second moment or covariance matrix is directly related to computing the SVD decomposition of $\mathbf{X}$ or $\widetilde{\mathbf{X}}$, with the eigenvalues being related to the singular values $\sigma_i$ of the corresponding matrices as $\lambda^u_i= \sigma^2_i(\mathbf{X})/N_r$ or $\lambda^c_i= \sigma^2_i(\widetilde{\mathbf{X}})/N_r$, 
 respectively, and the eigenvectors being the right singular vectors of the corresponding data matrix. 

\paragraph*{PCA spectrum} ---
Some of the quantities of primary interest in 
PCA are the normalised eigenvalues $\pi^{c/u}_i$ of the aforementioned matrices, i.e.,
\begin{equation}
\pi^{c/u}_i=\frac{\lambda^{c/u}_i}{\sum^{\min(N_r,m)}_{i=1}\lambda^{c/u}_i}.
\end{equation}
Since both the covariance matrix and the second-moment matrix are positive semidefinite, the quantities  $\{\pi^{c/u}_i\}$ can be formally regarded as probability distributions, as they satisfy the properties of being normalized and non-negative. 
In the case of the centered data $\widetilde{\mathbf{X}}$, the quantities $\pi^c_i$ represent the percentage of variance along the $i$-th principal eigenvector of the covariance matrix $\mathbf{\Sigma}$, and for this reason $\{\pi^c_i\}$ is referred to as the \textit{explained variance ratio}. A small explained variance along one eigendirection implies that the data are not much spread along that direction, and they can thus be regarded as lying on a 
lower-dimensional space. 

\paragraph*{PCA entropies} ---
The use of PCA as a compression algorithm involves a selection of $k<m$ eigenvectors associated with the $k$ largest normalized eigenvalues onto which the dataset is then projected. In many physical cases, however, there is no sharp separation between the magnitude of the various components, and the truncation cutoff is arbitrarily chosen case by case. 
A possible way to account for the full PCA spectrum is to introduce information-theoretic quantities such as entropies \cite{alter2000singular,varshavsky2006novel,varshavsky2007unsupervised}.
Given the normalized spectrum $\{\pi^{c/u}_i\}$ one can define Renyi-PCA entropies of order $n$ as 
\begin{equation}
    S^{c/u}_{{\rm PCA},n}=\frac{1}{1-n}  \log\Big(\sum_i (\pi^{c/u}_i)^n\Big),
    \label{eq:def-Reny}
\end{equation}
 which we call \textit{Renyi-PCA} entropies. All the Renyi-PCA entropies satisfy 
 \begin{equation}
     S^{c/u}_{\rm{PCA},n} \leq \log(\min(N_r,m)),
 \end{equation}
 as they attain their maximum value on the uniform distribution. 
We refer to the particular case
$n\to 1$ as  PCA-entropy 
\begin{equation}
    S^{c/u}_{\rm{PCA}}=-\sum_{i} \pi_i^{c/u}\log \pi_i^{c/u}.
    \label{eq:PCA-ent-def}
\end{equation} 

 These quantities provide information about the compressibility of the data, as they can serve as a measure of the flatness of the PCA spectrum. For example, if the explained variance ratio is uniform, one expects that all the components are relevant to describe the dataset and thus the dataset cannot be compressed and visualized in terms of a lower-dimensional space.  
Moreover, it has recently been shown that these
quantities can be used to extract critical behavior both in equilibrium 
classical systems and disordered quantum systems \cite{panda2023non,vanoni2024analysis}.

\subsubsection{Datasets}

In our analysis, we examine both $\mathbf{\Sigma}$ and  $\mathbf{M}$
in the quasistationary state and as a function of time. 
Concerning the study of quasistationarity, we will call \textit{steady state dataset} the collection of data matrices storing configurations $\{\vec{x}^{(\mu)}\}^{N_r}_{\mu=1}$ taken at time $t_c=L^z$ and extracted from independent simulations. In this case, one has to analyze a set of data matrices that store configurations for a given $p$ and $L$. We call instead \textit{time-dependent dataset} the collection of data matrices for different times. In this case, data matrices for different $p$, $L$, and $t$, are analyzed. 

%%
%%
%%%%%%%%%%%%%%%%%%%%%%%%%%%%%%%%%%%%%%%%%%%%%%%%%%%%%%%%%%%
%%%%%%%%%%%%%%%%%%%%%%%%%%%%%%%%%%%%%%%%%%%%%%%%%%%%%%%%%%%
\begin{figure}[h]
\includegraphics[width=.45\textwidth]{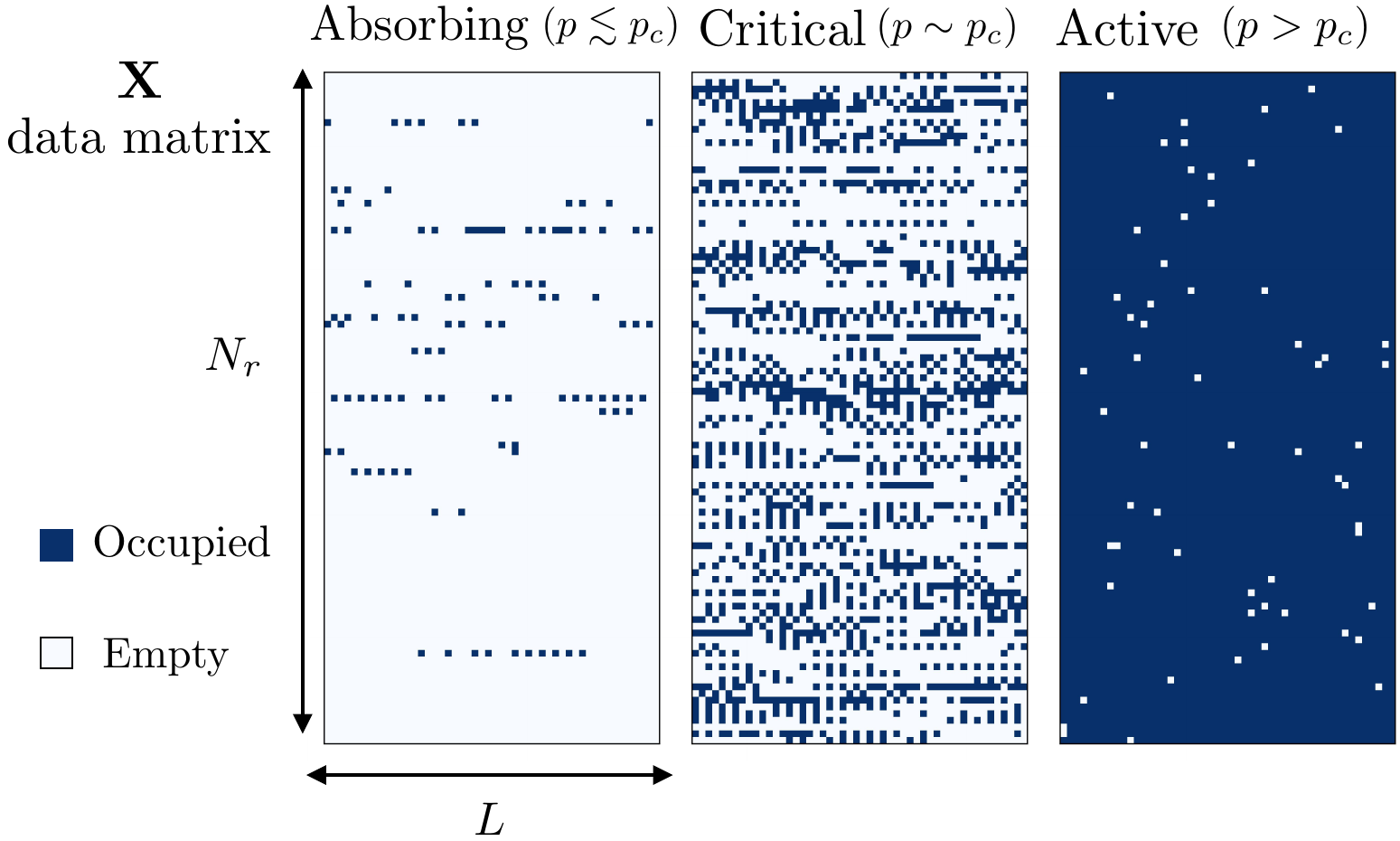}
\caption{Examples of data matrices $\mathbf{X}$ constructed from $N_r=100$ spatial configurations $\vec{x}^{(\mu)}$ of directed percolation on a lattice with  $m=L=50$, extracted at time $t_c$ from independent Monte Carlo simulations. These examples refer to three different values of $p$ which, from left to right, correspond to the absorbing, critical, and active phases for $L\to\infty$.
}
\label{fig:datamatrices}
\end{figure}
%%%%%%%%%%%%%%%%%%%%%%%%%%%%%%%%%%%%%%%%%%%%%%%%%%%%%%%%%%%
%%%%%%%%%%%%%%%%%%%%%%%%%%%%%%%%%%%%%%%%%%%%%%%%%%%%%%%%%%%
%%
%%
In all our analyses, the data matrices $\mathbf{X}$ 
and
$\widetilde{\mathbf{X}}$ have dimensions $N_r \times L$ and store the configurations of the specific model under study, obtained by evolving it from an initial active state $\mathcal{C}_{\rm{full}}=(1_1,...,1_L)$ in which all lattice sites are occupied. 
Examples of such data matrices are shown in Fig.~\ref{fig:datamatrices}. The observations $\vec{x}^{(\mu)}$ have dimension $m=L$ and each of them denotes a point on an $L$-dimensional unit hypercube. 
For all of our simulation data we choose $N_r>L$, which allows us to estimate second-moment matrix elements in a statistically accurate way \cite{potters2020first}. Choosing $N_r<L$ would not be statistically significant, as it is not possible to reconstruct faithfully $L^2$ correlations using $O(LN_r)<L^2$ data.

In order to provide a data-driven characterization of the samples, two approaches are possible. If one is interested in studying fluctuations of the data with respect to its mean, then centering  the data is the appropriate choice \cite{blum2020foundations}.
Most applications of PCA indeed involve a centering of the data $\mathbf{X}\mapsto \widetilde{\mathbf{X}}$ (see Eq.~\eqref{eq:centering}), followed by the analysis of the covariance matrix $\mathbf{\Sigma}$ which is directly obtained from $\tilde{\mathbf{X}}$. 
Connections between the properties of  PCA  of  centered and uncentered matrices are discussed in Ref.~\cite{cadima2009relationships}. 

In the present case, instead, we find that the data analysis obtained without centering the data is more suited to making qualitative and quantitative statements. 
As noted in the literature~\cite{Jolliffe}, this type of analysis can be relevant if the data are such that the origin is an important point of reference. This is in fact the case in our study, since the origin, for both models under study, represents their absorbing state configuration, namely $\mathcal{C}_\mathrm{vac}=(0_1, \dots, 0_L)$. 
Hence, here it is physically meaningful to study the variation of the data about this point. In fact, the emergence of a finite spatial density of particles in the system corresponds to an average displacement of the data points from the origin. Accordingly, the center of mass of the data $\mathbf{X}$ contains important information about the phase in which the system is. 
In our specific case, the data points $\vec{x}^{(\mu)}$ lie on the vertices of an $L$-dimensional unit hypercube, given that $ x_i^{(\mu)} \in\{0,1\}$ with $i=1, \ldots, L$. 
As the probability $p$ is increased (see Sec.~\ref{sec:models}), one expects the data to move from being close to the origin --- which corresponds to the absorbing state --- to being displaced towards the corner which is the farthest from it --- corresponding to the fully occupied active state.
The displacement of the centroid of the data from the origin is therefore related to the presence of an active phase. 

A natural question that arises is whether the spectral properties of the covariance matrix $\mathbf{\Sigma}$ or of the second-moment matrix $\mathbf{M}$, which are naturally expected to be sensitive to such a change, can be used to characterize the kind of transition undergone by the system as $p$ is varied. In our examination below, we find that both kinds of analyses can be used to describe the presence of such a phase transition, with an explicit connection to physical observables in the uncentered case. 
In view of this connection,
below we focus on the analysis of $\mathbf{M}$, while we present the results concerning $\mathbf{\Sigma}$ in Appendix~\ref{app:Centered}. Let us note, in addition, that also a centered analysis based on the quantified principal components, defined and discussed in  Appendix~\ref{app:Complementary} for DBP, can be physically informative.

\section{Results for directed bond percolation}\label{sec:PCA_DP}

\subsection{Largest normalized eigenvalue}\label{sec:PCA_DP_largest}

When characterizing a phase transition, the first step is to determine the critical point. In the case of directed percolation, the critical value $p_c$ of the probability $p$ can be determined by looking at the temporal evolution of the particle density $\rho (t)$, starting from a fully active configuration. In fact, at the critical point $p=p_c$, it  is expected to decay algebraically to zero as in Eq.~\eqref{eq:decay_rho}. In the absorbing regime, instead, an exponential decay, $\rho (t) \sim e^{-t/\xi_\parallel}$, is expected, while in the active phase $\rho(t)$ approaches a non-vanishing quasistationary plateau value \cite{marro2005nonequilibrium}.

\begin{figure}[h]
{\includegraphics[width=.45\textwidth]{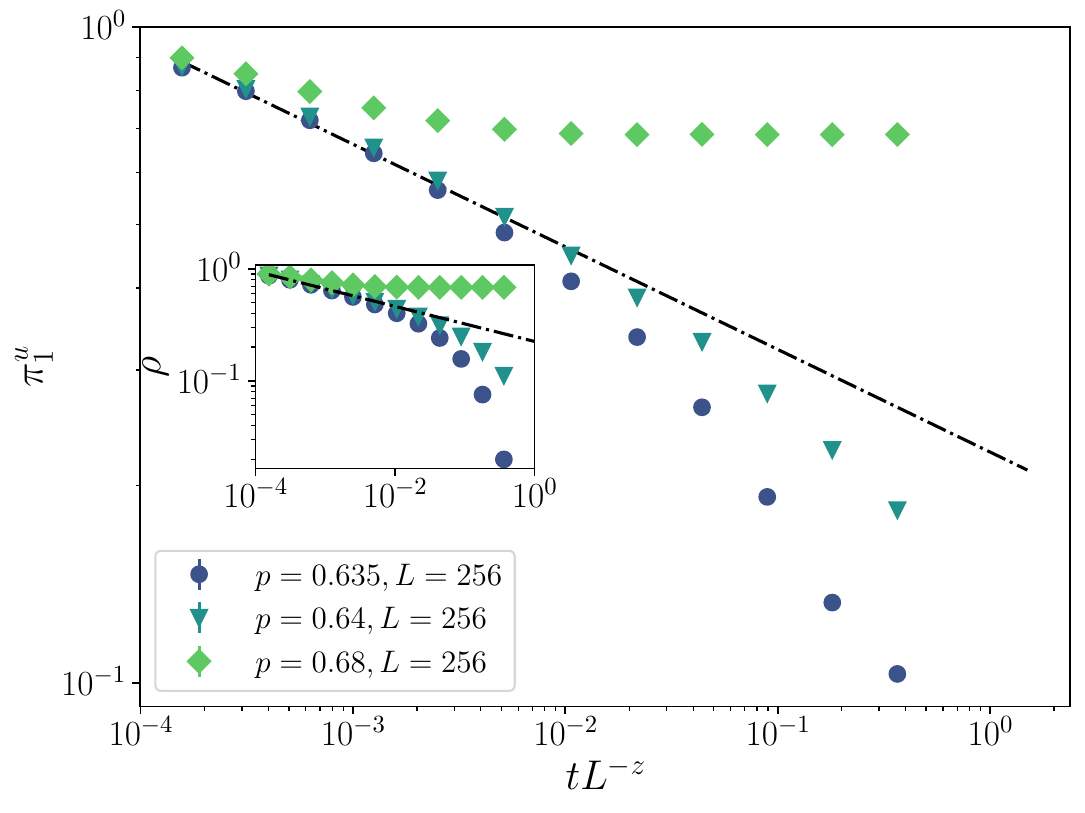}}
\caption{Largest normalized eigenvalue $\pi^u_1$ for DBP, as a function of time, obtained from the uncentered PCA analysis with $N_r=3000$ and for various values of $p$. Points are averaged over 5 repetitions. The standard deviation of the points is smaller than the dimension of the markers.  
The inset shows the corresponding density $\rho$ of active particles. The dotted-dashed line is a guide to the eye and corresponds to $\alpha = 0.157$ (see Eq.~\eqref{eq:decay_rho}).
}
\label{fig:pi1u_bendings}
\end{figure}

In Fig.~\ref{fig:pi1u_bendings}, we verify whether a similar behavior holds for the largest normalized eigenvalue $\pi^u_1$. In particular, we show three curves for $\pi^u_1(t)$ obtained for values of $p$ close to $p_c$.  We find that the critical point $p_c$ can be approximately determined by checking when the longest-lived algebraic decay occurs ($p=0.64$). In addition, the decay exponent for $\pi^u_1(t)$ at the critical point is in a reasonable agreement, considering the modest system size and the value of $p$,  with the exponent $\alpha$ which controls the decay of the density (and which was determined via accurate Monte Carlo simulations, see Tab.~\ref{table:tabexpDPPC}). We emphasize that this basic factual observation is actually the starting point for the more refined analysis we present below. This is a first clear indication that the PCA spectrum already carries considerable information about the system.
%%
%%
%%%%%%%%%%%%%%%%%%%%%%%%%%%%%%%%%%%%%%%%%%%%%%%%%%%%%%%%%%%%
%%%%%%%%%%%%%%%%%%%%%%%%%%%%%%%%%%%%%%%%%%%%%%%%%%%%%%%%%%%%
%%%%%%%%%%%%%%%%%%%%%%%%%%%%%%%%%%%%%%%%%%%%%%%%%%%%%%%%%%%%
%%%%%%%%%%%%%%%%%%%%%%%%%%%%%%%%%%%%%%%%%%%%%%%%%%%%%%%%%%%%
%%
%%

\begin{figure}[b]
{\includegraphics[width=.45\textwidth]{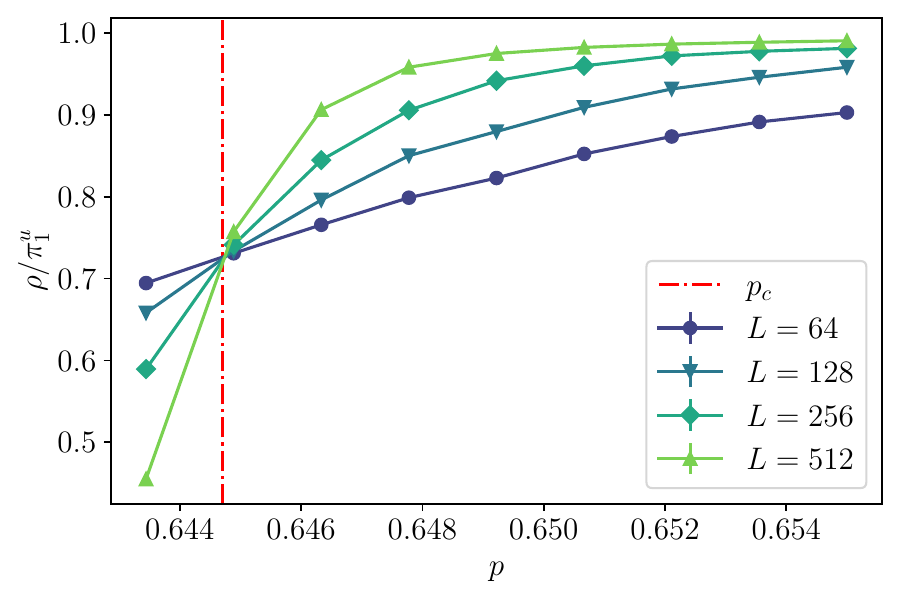}}
\caption{Behavior of the ratio $\rho/\pi^u_1$ for DBP. While above the critical point, the two quantities show an extremely good agreement, below the critical point the first uncentered normalized eigenvalue has a smaller overlap with the density. Each point is obtained using $N_r=2000$ and averaging over 5 repetitions.
}
\label{fig:DP_rho_pi1_u_ratio}
\end{figure}

We then examine if and how the steady-state values of the largest normalized eigenvalue $\pi^u_1$  and the particle density $\rho$ are related, by plotting their ratio $\rho/\pi^u_1$ as a function of $p$, shown in Fig.~\ref{fig:DP_rho_pi1_u_ratio} for various values of the system size $L$.
Upon increasing $L$, the ratio appears to converge to $\simeq 1$ in the active phase,
while we find that it decreases near the critical point and it keeps doing so upon decreasing
$p$ within the absorbing phase. 

\begin{figure}[h] 
     \includegraphics[scale=0.6]{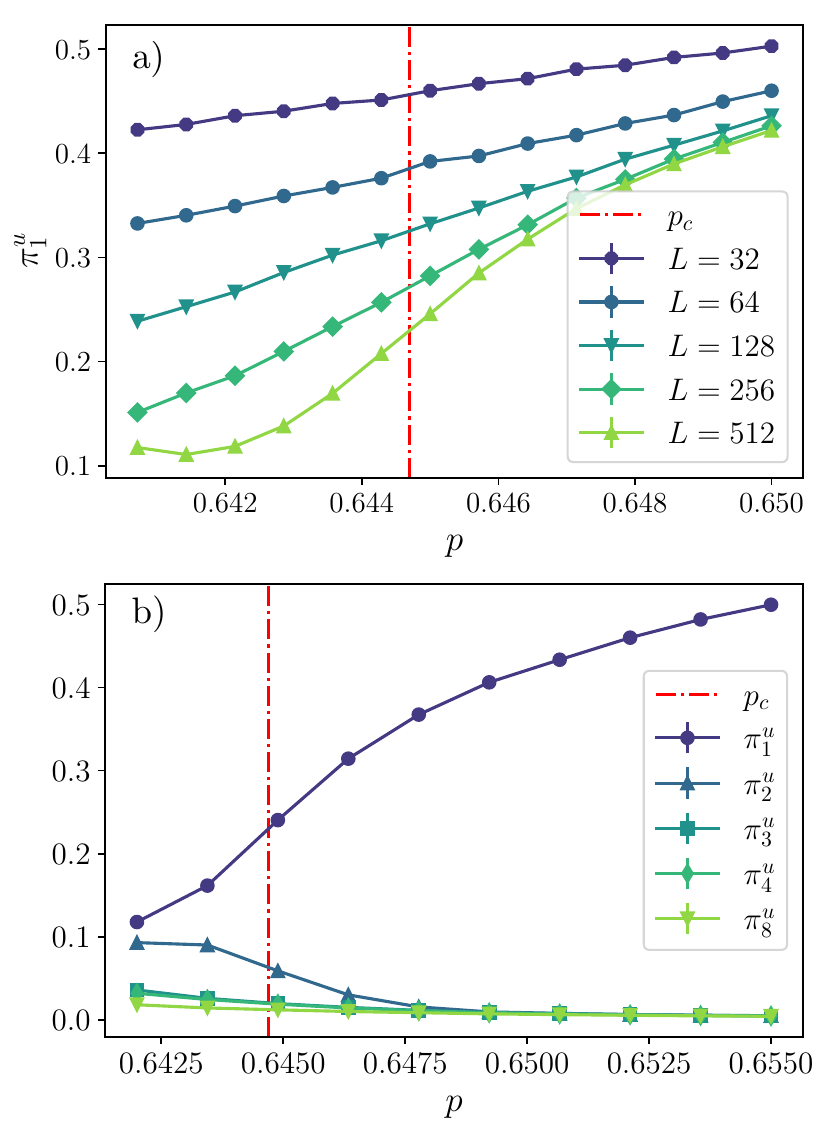}
    \caption{(a) Largest uncentered normalized eigenvalue $\pi^u_1$ in the vicinity of the critical point obtained for $N_r=3000$. Each data point is averaged over 5 realizations. (b) Behavior of the first few largest normalized eigenvalues for DBP close to the critical point for $L=512, N_r=2000$. Each point is averaged over 5 repetitions. Upon increasing the probability $p$ towards the active phase most weight of the normalized spectrum is contained in the first principal component. At the critical point $p = p_c$, the system still shows a clear separation of magnitude between the first component, i.e., the largest eigenvalue, and the others.}
    \label{fig:DP_Steady_Components}
\end{figure}

Panel (a) of Fig.~\ref{fig:DP_Steady_Components} shows the dependence of the stationary value of $\pi^u_1$ on the probability $p$ close to the transition point (vertical dash-dotted line), for various values of the system size $L$. Upon increasing $L$, the value of $\pi^u_1$ shows a clear inflection close to $p_c$, as one would expect for the order parameter of a continuous phase transition.
For the largest value of $L$ considered in panel (a), Fig.~\ref{fig:DP_Steady_Components}(b) shows the dependence on $p$ of the largest four eigenvalues $\pi^u_{1,2,3,4}$ and of $\pi^u_8$ close to the transition point. This shows that, in the active phase, the largest eigenvalue is actually significantly larger than the others, such that the ratio $\rho/\pi^u_1$ tends to one in the active phase. 

%%
%%
%%%%%%%%%%%%%%%%%%%%%%%%%%%%%%%%%%%%%%%%%%%%%%%%%%%%%%%%%%%
%%%%%%%%%%%%%%%%%%%%%%%%%%%%%%%%%%%%%%%%%%%%%%%%%%%%%%%%%%%
%%%%%%%%%%%%%%%%%%%%%%%%%%%%%%%%%%%%%%%%%%%%%%%%%%%%%%%%%%%
%%%%%%%%%%%%%%%%%%%%%%%%%%%%%%%%%%%%%%%%%%%%%%%%%%%%%%%%%%%
%%
%%
%%
%%
%%%%%%%%%%%%%%%%%%%%%%%%%%%%%%%%%%%%%%%%%%%%%%%%%%%%%%%%%%%
%%%%%%%%%%%%%%%%%%%%%%%%%%%%%%%%%%%%%%%%%%%%%%%%%%%%%%%%%%%
%%%%%%%%%%%%%%%%%%%%%%%%%%%%%%%%%%%%%%%%%%%%%%%%%%%%%%%%%%%
%%%%%%%%%%%%%%%%%%%%%%%%%%%%%%%%%%%%%%%%%%%%%%%%%%%%%%%%%%%
%%%%%%%%%%%%%%%%%%%%%%%%%%%%%%%%%%%%%%%%%%%%%%%%%%%%%%%%%%%
%%%%%%%%%%%%%%%%%%%%%%%%%%%%%%%%%%%%%%%%%%%%%%%%%%%%%%%%%%%
\begin{figure}[h]
    \includegraphics[scale=0.55]{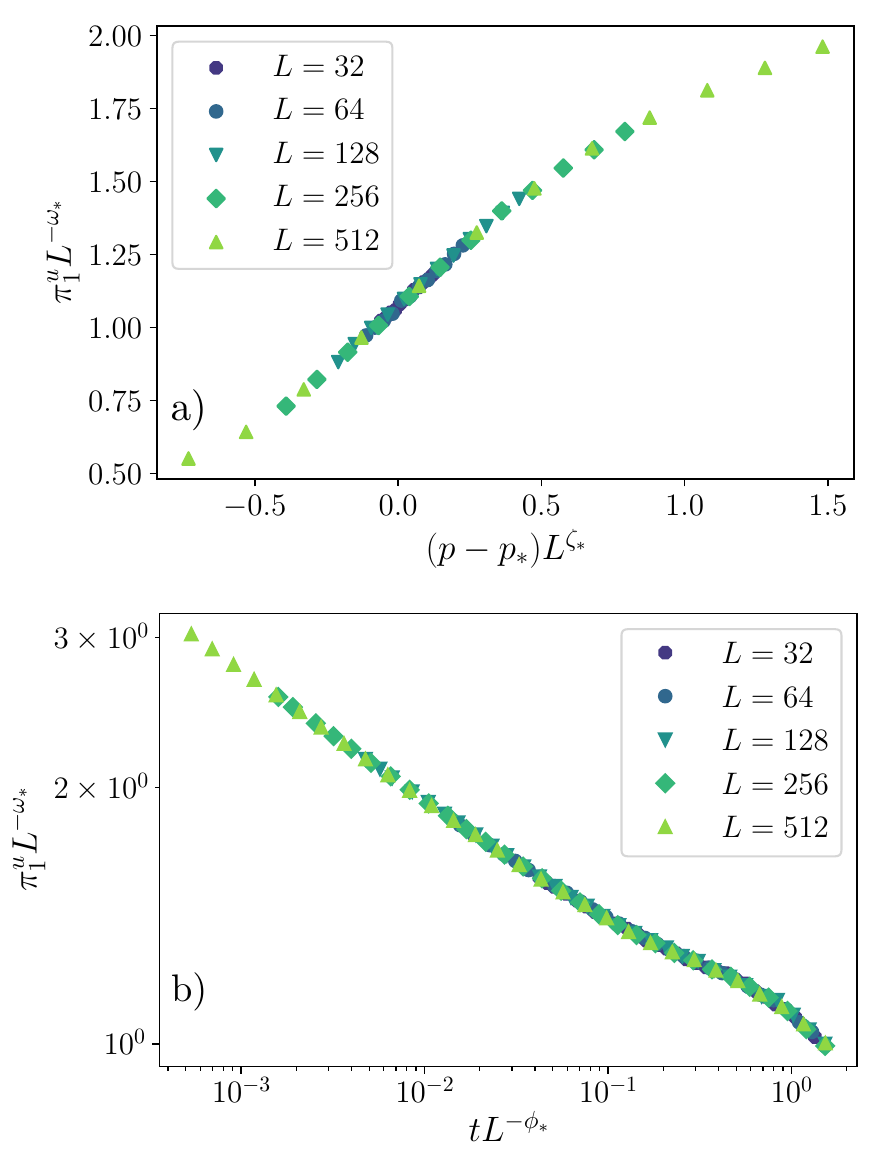}  
    \caption{(a) Finite-size scaling of $\pi^u_1$ in the steady state. The values for which the best data collapse is obtained are $p_*=0.64474(5)$, $\zeta_*=0.904(4)$, $\omega_*=-0.246(3)$. (b) Time-dependent finite size scaling of $\pi^u_1$. The estimated values are $\phi_*=1.58(2)$, $\omega_*=-0.248(3)$. In both cases, data were obtained for $N_r=3000$ and each data point is a result of an average over 5 realizations.
    }
    \label{fig:DP_Finite_Size_Scalings}
\end{figure}
%%%%%%%%%%%%%%%%%%%%%%%%%%%%%%%%%%%%%%%%%%%%%%%%%%%%%%%%%%%
%%%%%%%%%%%%%%%%%%%%%%%%%%%%%%%%%%%%%%%%%%%%%%%%%%%%%%%%%%%
%
%
As we show below, $\pi^u_1$ can be used to extract some universal features of the transition. 
In particular, we perform a finite-size scaling analysis of $\pi^u_1$ obtained from a PCA based on the steady-state configurations of DBP.
We assume the finite-size scaling form 
\begin{equation}\label{eq:FSS_steady}
    \pi^u_1(p,L)=L^{\omega}f((p-p_c)L^{\zeta}).
\end{equation}
If the direct relationship between $\pi^u_1$ and the particle density $\rho$ holds, 
the values of the exponents $\omega$ and $\zeta$ are expected to match those which result in a corresponding scaling collapse of the density $\rho$. These exponents are given by $\omega = -\beta/\nu_\perp \approx -0.252$ and $\zeta =  1/\nu_{\perp} \approx 0.911 $. 
To perform the finite-size scaling analysis, we adopt the method of Ref.~\cite{kawashima1993critical}, which is described in Appendix \ref{app:FSS}. We denote by $p_*$, $\omega_*$, and $\zeta_*$ the values of $p_c$, $\omega$, and $\zeta$ which lead to the best data collapse on a master curve $f$.

In Fig.~\ref{fig:DP_Finite_Size_Scalings}, we report the finite-size scaling analysis performed on the basis of the lattice configurations in the steady state, finding good agreement of $\omega_*=-0.246(3)$  and $\zeta_*=0.904(4)$ with the values $\omega$ and $\zeta$ of the DP universality class reported above. This fact indicates that, although $\pi^u_1$ is not the particle density, it captures the critical properties in a similar manner.
A way to extract further information on the critical exponents is to perform a finite-size scaling of the time dependence of $\pi^u_1$ for $p= p_c$, i.e., during its relaxation towards the vanishing stationary value. In this case, 
we use the scaling form  
\begin{equation}\label{eq:FSS_time}
    \pi^u_1(L,t)=L^{\omega}g(tL^{-\phi}),
\end{equation}
where, still assuming that $\pi^u_1$ behaves as the particle density, $\omega$ is given as discussed after Eq.~\eqref{eq:FSS_steady}  and $\phi = z =1.580$ in terms of the exponents in Tab.~\ref{table:tabexpDPPC}. 
Figure \ref{fig:DP_Finite_Size_Scalings} shows the result of the finite-size scaling. Also in this case, values $\omega_*=-0.248(3)$ and $\phi_*=1.58(2)$ determined from the best data collapse are fairly close to 
the values expected for the DP universality class.

\subsection{Analytical relation between principal components and order parameter}

The similarity in the scaling properties of the largest normalized eigenvalue $\pi^u_1$ and of the density $\rho$, illustrated in Fig.~\ref{fig:DP_rho_pi1_u_ratio}, can be understood through considerations on the SVD of the data matrix $\mathbf{X}$. Given an $n\times n'$ matrix $A$, we can perform its SVD decomposition and write $A=UDV^T$,  where $D$ is an $r\times r$ diagonal matrix, with $r=\text{rank}(A)$, having the singular values $\sigma_i(A)$ ($i = 1, \ldots, r$) along the diagonal, while $U$ and $V^T$ store the singular vectors $\mathbf{u}_i$ and $\mathbf{v}^T_i$, respectively, of $A$. 
In terms of these singular vectors, the SVD decomposition of $A$ can be written as 
\begin{equation}
    A=\sum^r_{i=1}\sigma_i(A) \mathbf{u}_i\mathbf{v}^T_i.
\end{equation}
We can define the truncation $A_k$ of $A$ to the first $k$  largest singular values (with $k<r$) as 
\begin{equation}
    A_k=\sum^k_{i=1} \sigma_i(A)\mathbf{u}_i\mathbf{v}^T_i,
\end{equation}
where the matrix $A_k$ has rank $k$.  It is possible to show that this decomposition is optimal
among all the possible rank-$k$ approximations \cite{eckart1936approximation, blum2020foundations}. This means that, for any matrix $B$ of rank at most $k$, the following inequality holds:
\begin{equation}
    \|A-A_k\|_F\leq \|A-B\|_F,
\end{equation}
where the Frobenius norm $\|\ldots\|_F$ of a matrix 
 $A$ is defined as $\|A\|_F=\sum_{ij}A^2_{ij}$.
Accordingly, the error made by approximating $A$ with $A_k$ is smaller than the error that one would make by approximating it with any other matrix of rank at most $k$. In this respect, $A_k$ is said to be the best rank-$k$ approximation of $A$.
Note that the sum of the squared singular values $\sigma_i^2(A)$ of a matrix $A$ equals
the Frobenius norm $\|A\|_F$ of the matrix \cite{blum2020foundations} (see, e.g., 
Ref.~\cite{trefethen2022numerical} for a proof).
For a binary dataset $X_{\mu i}=x^{(\mu)}_i \in \{0,1\}$, such as the one considered in the present work, one  has $X^2_{\mu i}=X_{\mu i}$ and thus
    \begin{equation}
       \|\mathbf{X}\|_F=\sum^{N_r}_{\mu=1}\sum^{L}_{i=1}X^2_{\mu i}= \sum^{N_r}_{\mu=1}\sum^{L}_{i=1}X_{\mu i}=N_r L \rho, 
    \end{equation}
    where in the last equality we used the definition of the density.
This equality implies that 
    \begin{equation}
        \sum^r_{i=1}\sigma^2_i(\mathbf{X})= N_r L \rho ,
    \end{equation}
or, equivalently, 
     \begin{equation} \label{eq:sumlambdaequalrho}
        \sum^r_{i=1}\lambda^u_i= L \rho ,
    \end{equation}
where $\lambda^u_i$ are the unnormalized eigenvalues of the second moment matrix $\mathbf{M}$, which are related to $\sigma^2_i$ as discussed after Eq.~\eqref{eq:def-M}.
     
    Equation (\ref{eq:sumlambdaequalrho}) expresses the equality the trace of $\mathbf{M}$ in the original basis and in the basis which diagonalizes $\mathbf{M}$.
    When a clear difference in the order of magnitude emerges between the first (largest) eigenvalue and the remaining ones, one can neglect the subleading contributions given by the latter to 
    Eq.~\eqref{eq:sumlambdaequalrho}, finding that $\lambda^u_1/L\simeq \rho $. This is precisely what is observed deep within the active phase where, as $p$ increases, the rank of $\mathbf{X}$ tends to be one and most of the weight tends to be contained in the first principal component. 
    Although Eq.~(\ref{eq:sumlambdaequalrho}) involves the largest unnormalized component $\lambda_1^u$, our numerical result show that also $\pi^u_1$ and $\rho$ are related.
    In particular, deep in the active phase where $\lambda^u_1 \gg\lambda^u_{i>1}$, $\pi^u_1$ and $\rho$ have the strong overlap displayed in Fig.~\ref{fig:DP_rho_pi1_u_ratio}, while at the critical point, the relationship between these two quantities is weakened by the fact that part of the spectral weight is due to other components. 
    Nevertheless, also close to the critical point, the largest uncentered explained variance $\pi^u_1$ shows a scaling behavior compatible with that one of the density. Keeping only the truncation of $\mathbf{X}$ to the largest uncentered component is what is called the \textit{best rank-1} truncation. The numerical data and analytic arguments presented above therefore suggest that the best rank-1 truncation of the data is already able to capture and reproduce the critical behavior of DBP.

\subsection{PCA entropies} 
\label{ss:entr-DP}

Entropies can also play a role in the characterization of absorbing state phase transitions \cite{EntropyDP,EntropySIS}. 
It is natural to wonder whether information-theoretic quantities, such as the Renyi-PCA entropies defined in Sec.~\ref{sec:methods}, which
involve the entire PCA spectrum beyond the largest eigenvalues considered above can be sensitive to critical behavior. 

\begin{figure}[t]
{\includegraphics[width=.45\textwidth]{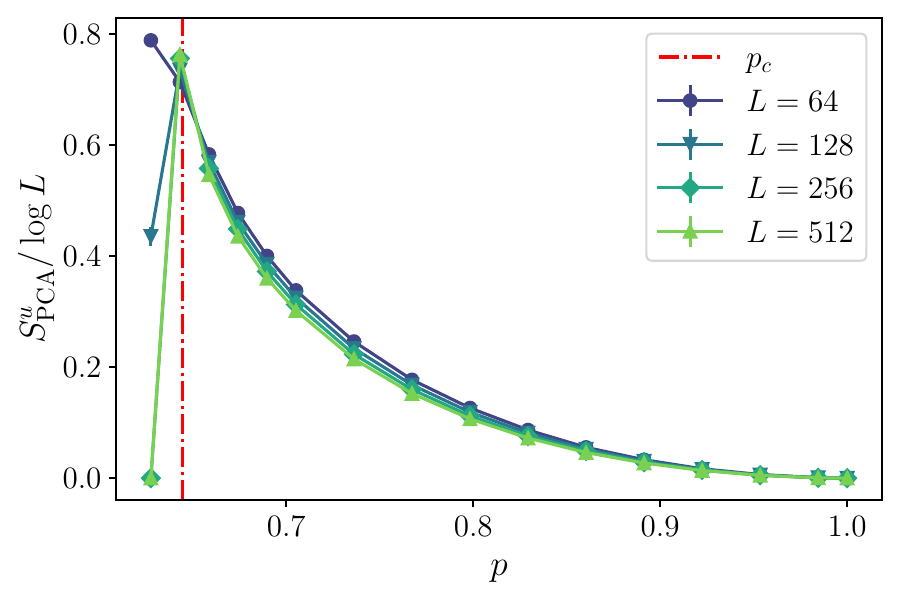}}
\caption{Normalized uncentered PCA entropy. As the system size $L$ increases, the entropy shows a transition in the vicinity of the critical point. Data points are obtained using $N_r=2000$. Each data point is the result of an average over 5 repetitions. 
}
\label{fig:DP_PCA_Phasediag}
\end{figure}

In Fig.~\ref{fig:DP_PCA_Phasediag}, we show the steady-state uncentered PCA entropy defined in Eq.~\eqref{eq:PCA-ent-def} as a function of $p$, for system sizes $L\in[64, 512]$, normalized to the maximum value it can attain ($\log L$). 
This quantity displays a peak in the vicinity of the critical point, marked by the vertical dash-dotted line. In the limit of infinite system size, no points are expected for $p<p_c$, as no fluctuations occur in the absorbing phase, and the PCA entropy should be exactly zero. The same kind of behavior is also seen in all the Renyi entropies in Eq.~\eqref{eq:def-Reny}, as discussed in Appendix \ref{app:Renyi-PCA entropies}.

\begin{figure}[b]
{\includegraphics[width=.45\textwidth]{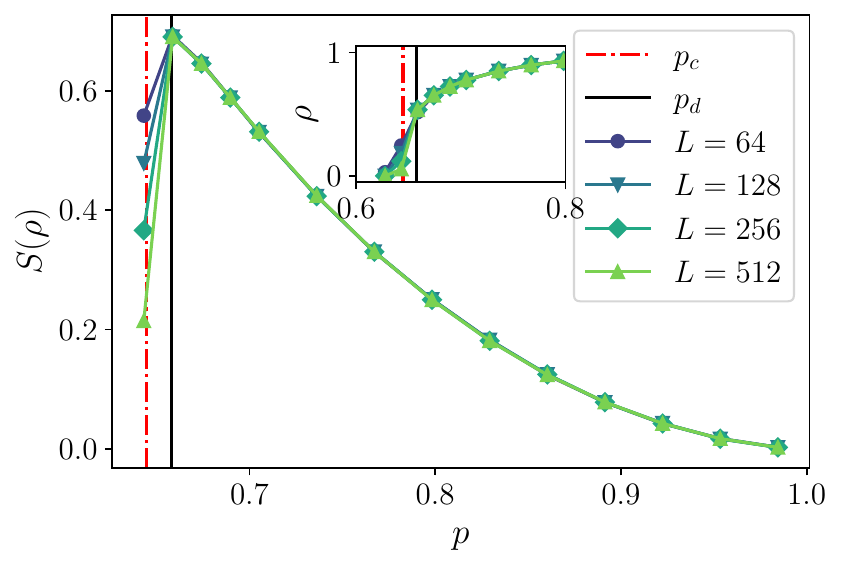}}
\caption{Shannon entropy of the order parameter, obtained for $N_r=2000$. Each data point is the result of an average over 5 repetitions. The inset shows the density $\rho$ of the active particles as a function of $p$. The peak of the Shannon entropy corresponds to the value $p_d$ of $p$, where the density attains the value 1/2.
}
\label{fig:DP_Shannon}
\end{figure}

In order to better understand the nature of the PCA entropy, we want to compare it with other commonly used measures of information in statistical mechanics. The most suitable comparison would be the one with the true Shannon entropy $S(\mathbf{X})$ of the dataset, which is inaccessible, as the entries of $\mathbf{X}$ are correlated and the probability distribution of $\mathbf{X}$ is unknown at any time, even in the stationary state. We note that, under specific circumstances, PCA and Shannon entropies may be closely related~\cite{vanoni2024analysis}, although here it is not the case. In order to have another candidate we compare the PCA entropy with the Shannon entropy of the order parameter $\rho$, namely,
\begin{equation}
\label{eq:SE-rho}
    S(\rho)=-\rho \log \rho -(1-\rho ) \log(1-\rho). 
\end{equation}
The reason to consider this object is that, within a mean-field approximation, treating the variables $x^{(\mu)}_i$ as uncorrelated, with $x^{(\mu)}_i =1$ with probability $\rho(p)$ and 
$x^{(\mu)}_i =0$ with 
probability $1-\rho(p)$, 
one would naturally assign to the uncentered dataset $\mathbf{X}$ an entropy 
$S(\mathbf{X})=N_rLS(\rho)$. In the present case, it turns out that the Shannon entropy displays a peak at a value $p_d\approx 0.658 > p_c$  of the probability $p$, as shown in Fig.~\ref{fig:DP_Shannon}. Correspondingly, the density $\rho$ takes the value  $\rho=1/2$. 
Comparing the behavior of the PCA and the Shannon entropy, we conclude that the former features a strong non-mean-field character.
%
%
%%
%%
%%%%%%%%%%%%%%%%%%%%%%%%%%%%%%%%%%%%%%%%%%%%%%%%%%%
%%%%%%%%%%%%%%%%%%%%%%%%%%%%%%%%%%%%%%%%%%%%%%%%%%%
%%%%%%%%%%%%%%%%%%%%%%%%%%%%%%%%%%%%%%%%%%%%%%%%%%%
%%%%%%%%%%%%%%%%%%%%%%%%%%%%%%%%%%%%%%%%%%%%%%%%%%%
%%
%%
%%%%%%%%%%%%%%%%%%%%%%%%%%%%%%%%%%%%%%%%%%%%%%%%%%%
%%%%%%%%%%%%%%%%%%%%%%%%%%%%%%%%%%%%%%%%%%%%%%%%%%%
%%%%%%%%%%%%%%%%%%%%%%%%%%%%%%%%%%%%%%%%%%%%%%%%%%%
%%%%%%%%%%%%%%%%%%%%%%%%%%%%%%%%%%%%%%%%%%%%%%%%%%%

\section{Results for branching and annihilating random walks}\label{sec:PCA_BARW}

In this section, we apply the analysis previously performed for directed bond percolation to the case of branching annihilating random walks with an even number of offspring, which display an absorbing phase transition belonging to the PC universality class. For parity-conserving models, however, particular care has to be used when analyzing the datasets. In fact, 
as pointed out in Ref.~\cite{jensen1994critical}, quasistationarity of observables is better observed by performing averages which take into account only the configurations which are not in the absorbing state. For this reason, after having computed the data matrices by storing the $N_r$ realizations of the dynamics of the model, we discard the configurations which have reached the absorbing state. All the analysis we present in this section is thus based exclusively on these surviving configurations, which means that the PCA is performed on matrices where inactive configurations are removed, and only active configurations are taken into account. In order to distinguish the density computed over surviving samples from the one defined in Eq.~\eqref{eq: def_density}, we denote the former as $\rho_S$. To simplify the notation, hereafter we do not add the subscript $S$ to the other quantities, but we understand that they are computed over the surviving samples.

\subsection{Largest normalized eigenvalue} \label{sec:PC_largest}

\begin{figure}
{\includegraphics[scale=0.45]{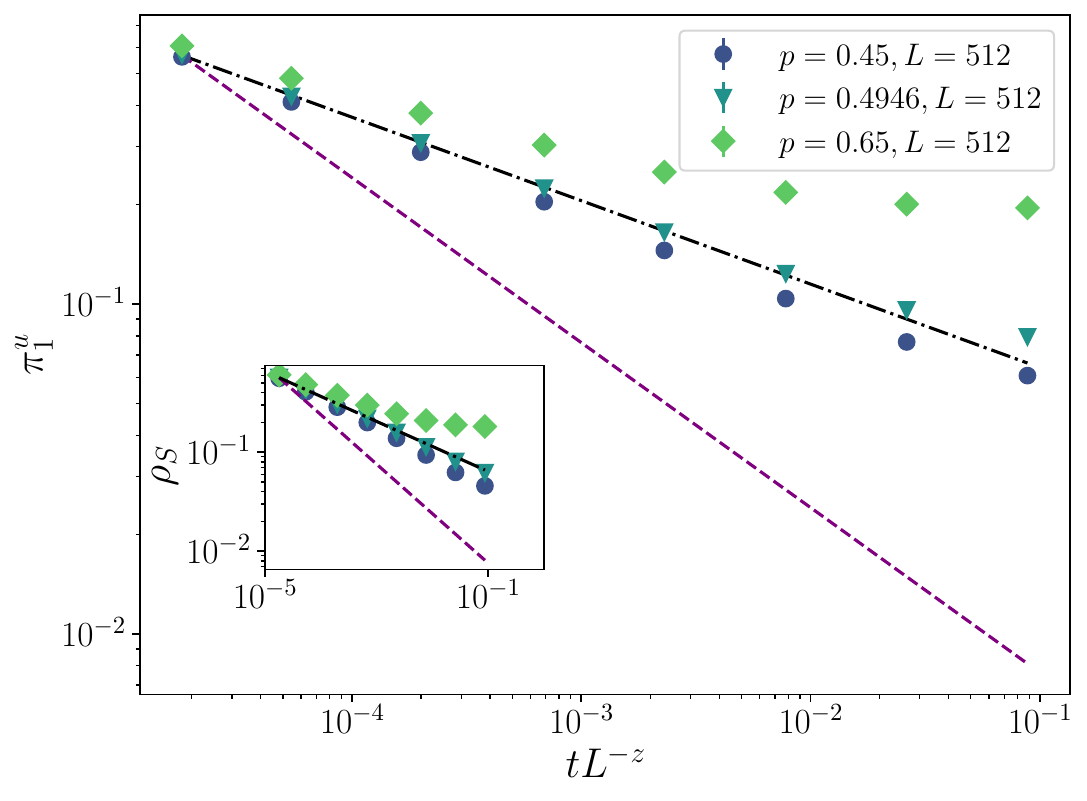}}
\caption{Largest normalized eigenvalue $\pi^u_1$ as a function of time, obtained from uncentered PCA analysis with $N_r=3000$ for BARW. Points are averaged over 5 repetitions.  The inset shows the corresponding density. The dotted-dashed line is obtained from a fit and corresponds to $\alpha = 0.253$, which deviates from the expected value $0.285(2)$ reported in Table \ref{table:tabexpDPPC}. The dashed line, instead, reports the expected decay for $p<p_c$, associated with $\alpha=1/2$. 
}
\label{fig:pi1u_bendings_PC}
\end{figure}

As we did for DBP, we first analyze the behavior of the largest normalized uncentered eigenvalue $\pi^u_1$ as a function of time, which is reported in Fig.~\ref{fig:pi1u_bendings_PC} for three values of $p$ corresponding to the active, critical, and absorbing phase and with $L=512$. 
In this case, the analysis is complicated by the fact that in parity-conserving models the decay of the density $\rho(t)$ as a function of time $t$ is algebraic also within the absorbing phase \cite{hinrichsen2000non}. Indeed, such algebraic decay is observed in Fig.~\ref{fig:pi1u_bendings_PC} both for $\pi^u_1$ and $\rho_S$.
The expected power-law decay in the absorbing phase is the one associated with the pair-annihilation fixed point in one dimension $\alpha=1/2$ (see Ref.~\cite{tauber2017phase,tauber2002dynamic}) and it is shown in Fig.~\ref{fig:pi1u_bendings_PC} as a dashed line. As a matter of fact, this power can be observed in numerical simulations only at very low values of $p$ or at very large system sizes and the decay exponent varies continuously for finite system sizes. 
For the sizes accessible to our numerical simulations (up to $L=512$),
it turned out to be difficult to disentangle finite-size effects from a possible crossover in the algebraic exponent. 
Indeed, the system sizes used in our simulations are much smaller than those used, for example, in Ref.~\cite{Park_2013}.  As a consequence, 
we did not manage to extract a decay exponent for $\pi^u_1(t)$ which is compatible with the value expected for $\alpha$ within PC universality class.
We notice that this is not a peculiarity of this case: such finite volume effects are typically observed in the presence of algebraic decay of correlations, and occur in equilibrium as well -- one example being Berezinski-Kosterlitz-Thouless phase transition.

%
%%
%%
%%%%%%%%%%%%%%%%%%%%%%%%%%%%%%%%%%%%%%%%%%%%%%%%
%%%%%%%%%%%%%%%%%%%%%%%%%%%%%%%%%%%%%%%%%%%%%%%%
%%%%%%%%%%%%%%%%%%%%%%%%%%%%%%%%%%%%%%%%%%%%%%%%
%%%%%%%%%%%%%%%%%%%%%%%%%%%%%%%%%%%%%%%%%%%%%%%%
%%
%%

\begin{figure}
   \includegraphics[scale=0.55]{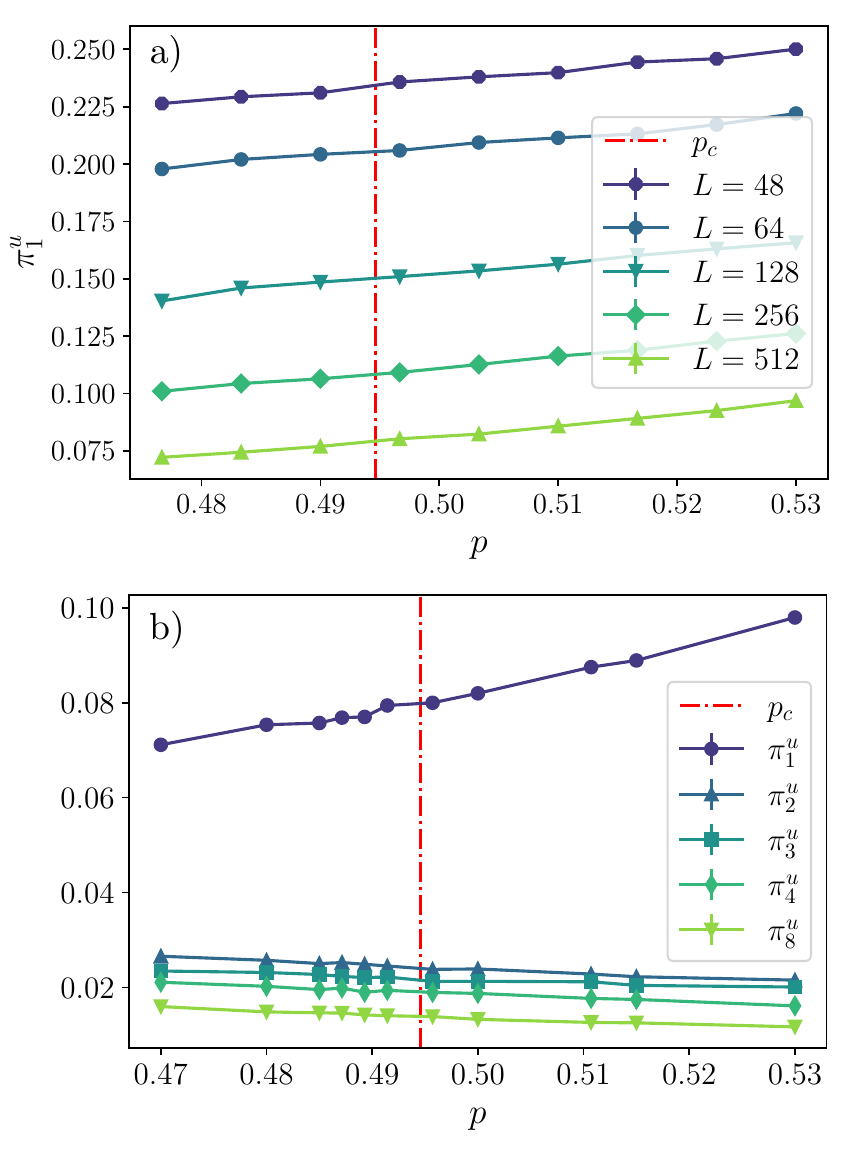}
    \caption{(a) Largest uncentered normalized eigenvalue $\pi^u_1$ for BARWe  as a function of the probability $p$, in the vicinity of the critical point $p\simeq p_c$ for various system sizes. Simulations were performed with $N_r=4000$, each point is averaged over 20 repetitions. (b) Dependence on $p$ of the first few largest components $\pi^u_{1,2,3,4,8}$ for BARWe, with $p$ close to the critical point $p_c$ and $L=512$, $N_r=3000$. Each point is averaged over 10 repetitions.
    }
    \label{fig:PC_Steady_Components}
\end{figure}

\begin{figure}
{\includegraphics[width=.45\textwidth]{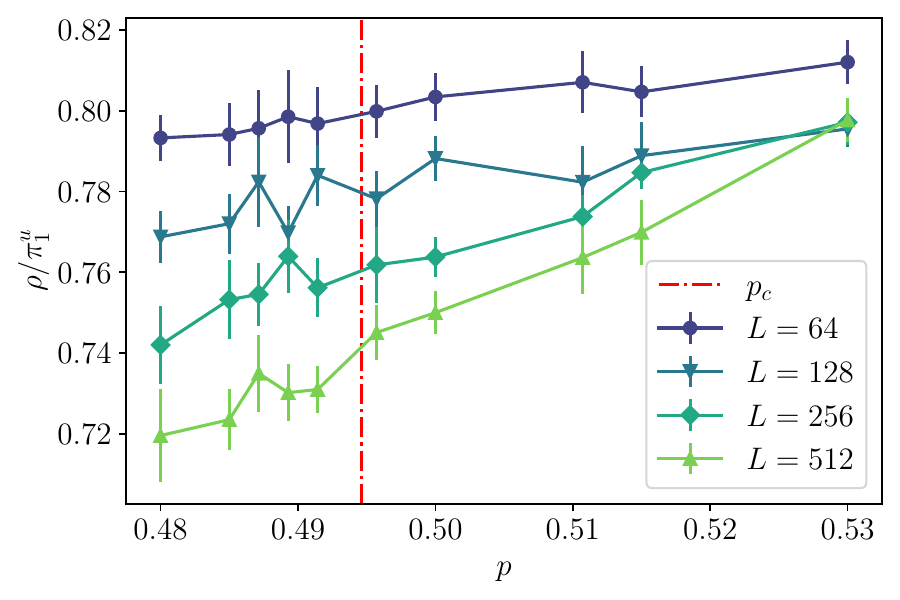}}
\caption{Dependence of the ratio $\rho/\pi^u_1$ for BARWe on the probability $p$, obtained for $N_r=3000$ and various values of the system size $L$. Each point is averaged over 10 realizations. }
\label{fig:PC_rho_pi1_u_ratio}
\end{figure}

In Fig.~\ref{fig:PC_Steady_Components}(a), we report the steady-state value of $\pi^u_1$ as a function of the probability $p$ close to the critical point $p\simeq p_c$. In this case, differently from DBP [see Fig.~\ref{fig:DP_Steady_Components}(a)], no net change in the normalized spectrum is observed, as shown in Fig.~\ref{fig:PC_Steady_Components}. 
In fact, as $p$ is varied, the largest normalized eigenvalue $\pi^u_1$ is not affected significantly as it displays a rather smooth behavior also upon increasing $L$, see Fig.~\ref{fig:PC_Steady_Components}(a). In order to inspect if the same occurs also to the subleading components, in Fig.~\ref{fig:PC_Steady_Components}(b) we plot the dependence on $p$ also of the subsequent eigenvalues $\pi^u_{2,3,4,8}$ for $L=512$.
Differently from DBP, and similarly to what is observed in panel (a) of the same figure, their behavior is still rather smooth upon increasing $p$ and, in addition, no clear quantitative separation emerges among the considered eigenvalues. For example, at the critical point, the first normalized eigenvalue $\pi^u_1$ is as large as the sum of $\pi^u_2$, $\pi^u_3$, and $\pi^u_4$. 
Accordingly, we expect that the direct relationship existing between the density and the normalized principal component in the case of DBP might be affected.
%
%
%%
%%
%%%%%%%%%%%%%%%%%%%%%%%%%%%%%%%%%%%%%%%%%%%%%%%%%%%%%
%%%%%%%%%%%%%%%%%%%%%%%%%%%%%%%%%%%%%%%%%%%%%%%%%%%%%
%%%%%%%%%%%%%%%%%%%%%%%%%%%%%%%%%%%%%%%%%%%%%%%%%%%%%
%%%%%%%%%%%%%%%%%%%%%%%%%%%%%%%%%%%%%%%%%%%%%%%%%%%%%
%%
%%
%
This ratio  displays a different behavior compared to what is observed for DBP, i.e., it does not suddenly approach $1$ as $p$ increases beyond $p_c$. For comparison, note that the plot in Fig.~\ref{fig:PC_rho_pi1_u_ratio} spans a significantly wider range of values of $p$ than that for DBP in Fig.~\ref{fig:DP_rho_pi1_u_ratio}.
%%
%%
%%%%%%%%%%%%%%%%%%%%%%%%%%%%%%%%%%%%%%%%%%%%%%%%
%%%%%%%%%%%%%%%%%%%%%%%%%%%%%%%%%%%%%%%%%%%%%%%%
%%%%%%%%%%%%%%%%%%%%%%%%%%%%%%%%%%%%%%%%%%%%%%%%
%%%%%%%%%%%%%%%%%%%%%%%%%%%%%%%%%%%%%%%%%%%%%%%%
%%
%%

\begin{figure}
    \includegraphics[scale=0.6]{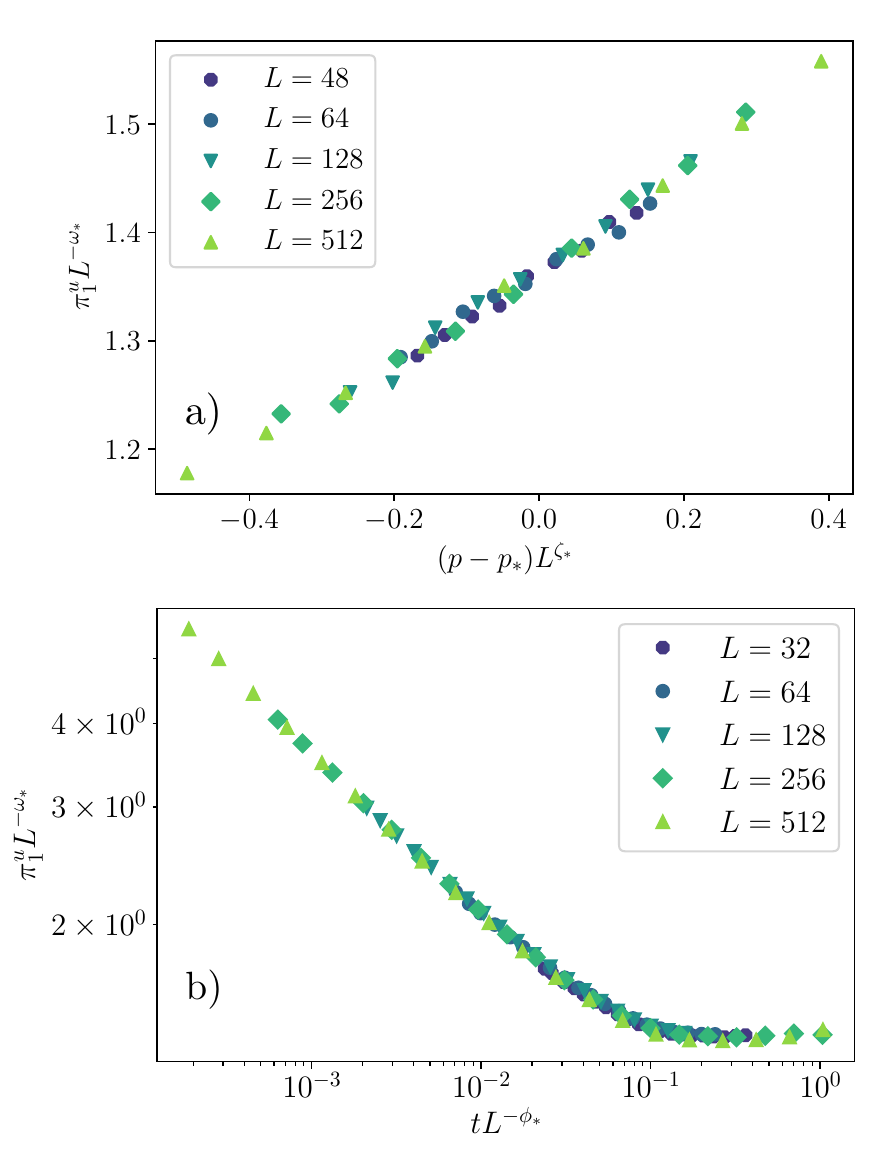}
    \caption{(a) Finite-size scaling of $\pi^u_1$ for BARWe in the steady state. The values $p_*$, $\zeta_*$ and $\omega_*$ of the corresponding parameters $p_c$, $\zeta$ and $\omega$ which yield the best data collapse according to Eq.~\eqref{eq:FSS_steady} are $p_*=0.500(9)$, $\zeta_*=0.45(5)$, $\omega_*=-0.45(1)$. (b) Finite-size scaling analysis  of the time dependence of $\pi^u_1(t)$ at the critical point $p=p_c$ (with $p_c$ given in Tab.~\ref{table:tabexpDPPC}).
    The values $\phi_*$ and $\omega_*$ of the corresponding parameters $\phi$ and $\omega$ which yield the best data collapse according to Eq.~\eqref{eq:FSS_time} are $\phi_*=1.74(3)$, $\omega_*=-0.459(3)$. 
    In both cases, simulations were performed for $N_r=4000$ and each data point is the result of averages over 20 realizations.
    }
    \label{fig:PC_Finite_Size_Scalings}
\end{figure}

In order to determine the critical exponents and the critical point, we perform a finite-size scaling analysis on the steady-state values of $\pi^u_1$ according to Eq.~\eqref{eq:FSS_steady}, with the result reported in Fig.~\ref{fig:PC_Finite_Size_Scalings}(a).
The estimate of the critical value $p_*$ of $p$ which yields the best data collapse is compatible with previous estimates \cite{zhong1995universality}, see also Tab.~\ref{table:tabexpDPPC}. However, differently from the case of DBP, the corresponding values of the exponents $\omega_*=-0.45(1)$ and $\zeta_* = 0.45(5)$ are not compatible with $-\beta/\nu_\bot=-0.50(2)$ and $1/\nu_\bot=0.54(2)$ (see Tab.~\ref{table:tabexpDPPC}), respectively.
As we argued previously, this discrepancy might be related to the different structures in the normalized spectrum, in which the subsequent principal components of the PCA do not differ by an order of magnitude from the largest one, see Fig.~\ref{fig:PC_Steady_Components}.

To proceed further with our analysis, we also perform a finite-size scaling  analysis of the time dependence of $\pi^u_1(t)$ during an evolution with $p=p_c$ (see Eq.~\eqref{eq:FSS_time}), reported in 
Fig.~\ref{fig:PC_Finite_Size_Scalings}(b). 
The value $\omega_*=-0.459(3)$ extracted from this analysis is compatible with the value extracted above from the analysis of the steady-state data and thus does not agree with the value $-\beta/\nu_{\perp}$ expected for PC. 
Concerning $\phi_*$, instead, we find a value $\phi_*=1.74(3)$ compatible with the dynamical critical exponent  $1.7415(5)$ of PC universality class (see Tab.~\ref{table:tabexpDPPC}). 

We conclude that, for this model, the similarity 
between the scaling properties of the particle density $\rho$ and the largest normalized eigenvalue $\pi_1^u$ is weakened by the fact that the PCA spectrum has a richer structure than in the case of DBP. Nevertheless, some quantities, such as the critical point $p_c$ and the dynamical critical exponent $z$, can be determined rather accurately with an analysis of the largest principal component. 

%%
%%
%%%%%%%%%%%%%%%%%%%%%%%%%%%%%%%%%%%%%%%%%%%%%%%%%%%%%%
%%%%%%%%%%%%%%%%%%%%%%%%%%%%%%%%%%%%%%%%%%%%%%%%%%%%%%
%%%%%%%%%%%%%%%%%%%%%%%%%%%%%%%%%%%%%%%%%%%%%%%%%%%%%%
%%%%%%%%%%%%%%%%%%%%%%%%%%%%%%%%%%%%%%%%%%%%%%%%%%%%%%
%%
%%

\subsection{PCA entropies}

\begin{figure}
{\includegraphics[width=.45\textwidth]{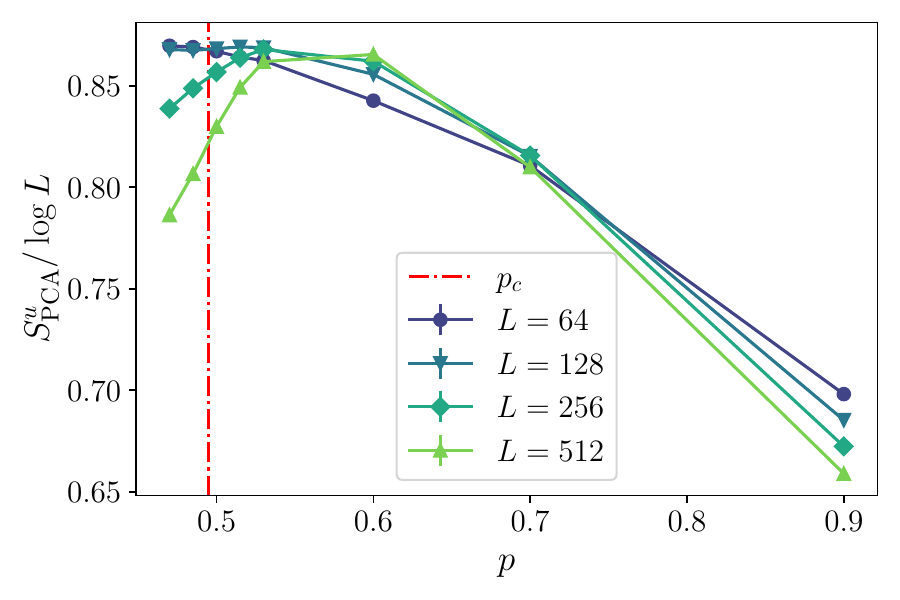}}
\caption{Normalized uncentered PCA entropy $S^u_{\rm PCA}$ for BARWe, normalized by $\log L$.
Simulations were performed with $N_r=3000$ each point is the result of an average over 10 repetitions.  }
\label{fig:PC_PCAu_Steady}
\end{figure}

\begin{figure}
{\includegraphics[width=.45\textwidth]{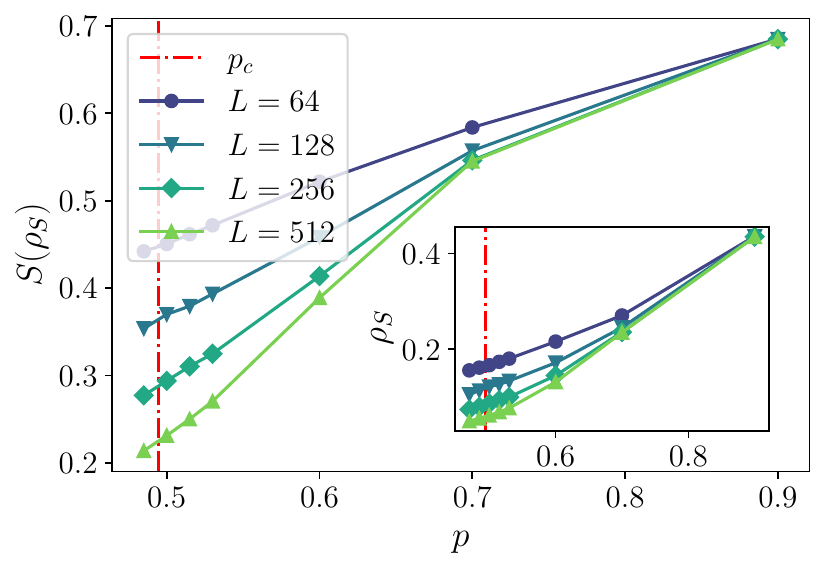}}
\caption{Shannon entropy of the order parameter as a function of $p$. For BARWe $S(\rho_S)$ peaks deep down in the active phase, as $\rho_S=0.5$ is deep far from the critical point.   Simulations were performed with $N_r=3000$ each point is the result of an average of over 10 repetitions.
}
\label{fig:PC_Shannon}
\end{figure}

In analogy with the analysis performed for DBP in Sec.~\ref{ss:entr-DP}, we analyze here the uncentered Renyi-PCA entropies in Eq.~\eqref{eq:def-Reny} for BARWe. Here we report only the results for the PCA entropy in Eq.~\eqref{eq:PCA-ent-def}, while those for the Renyi entropies are discussed in Appendix~\ref{app:Renyi-PCA entropies}.
The uncentered PCA entropy $S^u_{\rm PCA}$ in the stationary state of the model as a function of the probability $p$ is reported in Fig.~\ref{fig:PC_PCAu_Steady} for various system sizes $L$. The Shannon entropy $S(\rho_S)$ associated with the particle density $\rho_S$ (see Eq.~\eqref{eq:SE-rho}), instead, is reported in Fig.~\ref{fig:PC_Shannon}.
%
%%
%%
%%%%%%%%%%%%%%%%%%%%%%%%%%%%%%%%%%%%%%%%%%%%%%%%%%%%
%%%%%%%%%%%%%%%%%%%%%%%%%%%%%%%%%%%%%%%%%%%%%%%%%%%%
%%%%%%%%%%%%%%%%%%%%%%%%%%%%%%%%%%%%%%%%%%%%%%%%%%%%
%%%%%%%%%%%%%%%%%%%%%%%%%%%%%%%%%%%%%%%%%%%%%%%%%%%%
%%
%%
%
In this case, we find that for system size $L \in [64,512]$, the PCA entropy develops a peak at a value of $p$ which differs significantly from the critical value $p_c$ (vertical dash-dotted line). At the same time, however, the PCA entropy does not peak deep down in the active phase at $p\simeq 1$, in the point where the density is approximately $0.5$, where one would expect the peak the corresponding entropy of the dataset in mean-field approximation. 
%
%
%%
%%%%%%%%%%%%%%%%%%%%%%%%%%%%%%%%%%%%%%%%%%%%%%%%%%%%%%
%%%%%%%%%%%%%%%%%%%%%%%%%%%%%%%%%%%%%%%%%%%%%%%%%%%%%%
%%%%%%%%%%%%%%%%%%%%%%%%%%%%%%%%%%%%%%%%%%%%%%%%%%%%%%
%%%%%%%%%%%%%%%%%%%%%%%%%%%%%%%%%%%%%%%%%%%%%%%%%%%%%%
%%
%%

\section{Conclusions} 
\label{sec:conclusions}

In this work, we explored the use of principal component analysis (PCA) for studying nonequilibrium universality classes. We have shown that PCA can be successfully applied to determine the critical properties of the directed percolation universality class through the analysis of the largest normalized eigenvalue $\pi^u_1$ of the second moment matrix constructed from the configuration of the DBP model, either during its evolution or in its stationary state.

In this case, the success of this approach can be attributed to the fact that $\pi^u_1$, even close to criticality, is well separated from the 
smaller eigenvalues. 
At the same time, we have shown that entropic quantities that keep track of the entire PCA spectrum can serve as indicators of criticality, as shown in the case of DBP (Fig. \ref{fig:DP_PCA_Phasediag}). 
For models showing a PCA spectrum with a richer structure, such as BARWe, the similarity between the scaling properties of the largest normalized eigenvalue and of the order parameter of the model is considerably weaker --- a fact which we attribute to larger finite-volume effects. 
Nevertheless, some quantities, such as the critical point and the dynamical critical exponent, can still be inferred accurately from the analysis of the sole largest normalized eigenvalue $\pi^u_1$. 
Importantly, we elucidate our numerical observations by employing linear algebra arguments, which form the basis of PCA, thereby leveraging its interpretable nature.

Our findings show an interesting connection with low-rank approximations of datasets. In particular, our observations imply that low-rank (even rank-one) approximations of datasets can be successfully used to capture the universality of nonequilibrium phase transitions. 
This observation has important consequences for experiments since it implies that certain kinds of datasets can be compressed while still preserving universal properties. We note that similar conclusions were drawn concerning random graphs \cite{thibeault2024low}.

Some of our arguments hinge upon the binary nature of the datasets. Within this kind of dataset, it is plausible that our findings might apply not only to the models considered in this work, but to all models displaying a ``dominant'' principal component and a phase transition which can be characterized in terms of a uniform order parameter. It would be interesting to see if quantitative statements about rank-one approximations carry over to models with integer (but non-binary) or real variables.

The results presented here demonstrate that (linear) dimensional reduction works drastically different in different universality classes. Investigating how this occurs will likely require taking into account also non-linear effects. In this direction, possible approaches are provided by the kernel analysis, or even by a full data structure characterization based on network theory, which, so far, have been  only applied to models at equilibrium~\cite{panda2023non,sun2023network,mendes2023wave}.

\section*{Acknowledgments} 
We gratefully acknowledge R.~Andreoni, J.~Barbier, R.~Panda, C.~Vanoni and V.~Vitale for discussions and X.~Turkeshi for help in coding the finite-size scaling analysis. We thank U.~T{\"a}uber for discussions about the PC universality class. 
M.~D.~was partly supported by the MIUR Programme FARE (MEPH), by QUANTERA DYNAMITE PCI2022-132919, by the EU-Flagship programme Pasquans2, by the PRIN programme (project CoQuS) and, together with A.~G.~by the PNRR MUR project PE0000023-NQSTI.

\bibliography{main}

\appendix
\clearpage

\section{Finite-size Scaling}
\label{app:FSS}

As starting point for our discussion, let us consider a system at thermal equilibrium that undergoes a continuous phase transition upon tuning the value of a parameter $g$ to its critical value $g_c$. Accordingly, in the thermodynamic limit, the correlation length of the order parameter fluctuations within the system for $g\to g_c$ increases as  
\begin{equation}
    \xi \sim |g-g_c|^{-\nu},
    \label{eq:app-def-nu}
\end{equation}
where $\nu$ is the critical exponent associated with the universality class to which the system under consideration belongs. Now, let us consider a quantity $O$ which, in the thermodynamic limit, displays a singularity  of the form 
\begin{equation}
    O\sim |g-g_c|^{\gamma_O},
    \label{eq:app-sing-td}
\end{equation}
where $\gamma_O$ is an exponent depending on $O$. 
If the system has a finite spatial extent of typical size $L$, the singularities mentioned above are replaced by a regular behavior as a function of $g$ predicted by the finite-size scaling theory at equilibrium \cite{barber1983finite,cardy2012finite}.
In particular, for $g \to g_c$ and sufficiently large $L$, Eq.~\eqref{eq:app-sing-td} is replaced by 
\begin{equation}\label{eq:fss_form}
    O(g,L)= L^{-\gamma_O/\nu} \widetilde{O}((g-g_c)L^{1/\nu}), 
\end{equation}
where $\widetilde{O}$ is a scaling function independent of the system size $L$. 
This scaling ansatz actually provides a method for determining both the critical exponents and the critical point. 
In fact, if the critical point $g_c$ and the exponents $\gamma_O$ and $\nu$ associated with the transition are known, plotting $L^{\gamma_O/\nu}O(g,L)$ as a function of $(g-g_c)L^{1/\nu}$ should result in a collapse of the data concerning the dependence of $O$ on $g$ for various systems sizes $L$, because the scaling function $\widetilde{O}$ is independent of $L$ (at least at the leading order as $L$ grows). 
Accordingly, one can reverse the argument and determine the critical exponents and the critical point by treating  $\nu$, $\gamma_O$, and $g_c$ as fitting parameters and by estimating them with the corresponding values  $\nu_*$, $\gamma_*$ and $g_*$ which yield the best data collapse onto a single master curve. For example, this might amount to minimizing some distance between the scaled data. 
In order to perform the finite-size scaling we adopt the method described in Ref.~\cite{kawashima1993critical}. 
The same scaling form as Eq.~\eqref{eq:fss_form} and its generalizations can be used to describe steady-state scaling observables for nonequilibrium phase transitions \cite{lubeck2004universal}.  
For every scaling quantity $O$ depending on the parameter $p$ (see Sec.~\ref{sec:models}) and in a finite volume of size $L$, one can assume a scaling form
\begin{equation}
    O(p,L)=L^\omega\widetilde{O}((p-p_c)L^\zeta), 
    \label{eq:scaling_ansatz}
\end{equation}
where $\omega$ and $\zeta$ could, in principle, depend on the quantity $O$ taken into consideration, i.e., $\omega=\omega_O$ and $\zeta=\zeta_O$. 
However, in an ideal scenario in which the (stationary and dynamic) properties of the system are actually controlled by a well-defined and growing correlation length, the exponent $\zeta$ should actually be independent of the observable and be fixed to $1/\nu$ (where $\nu$ is defined in Eq.~\eqref{eq:app-def-nu}), especially when the quantity under study displays a favorable scaling behavior.

We now briefly describe the approach we used for obtaining and optimizing the data collapse in the finite-size scaling analysis, following Ref.~\cite{kawashima1993critical}. The method consists in minimizing a cost function which is computed on the basis of the raw numerical data for $O(p,L)$ and the associated error $\sigma_O(p,L)$, expressed in terms of the scaled variables $x \equiv (p-\bar{p}_c)L^{\bar{\zeta}}$, $y\equiv OL^{-\bar{\omega}}$ and scaled errors $\sigma=\sigma_OL^{-\bar{\omega}}$ where $\bar{\zeta}$, $\bar{\omega}$ are the candidate scaling exponents and $\bar{p}_c$ the candidate critical value of $p$. 
The scaled data and errors are computed for all the values $\{L_k\}^{N_L}_{k=1}$ of the system size $L$  and for all the values $\{p_j\}^{N_p}_{j=1}$ of the probability $p$ considered in the numerical study. 
The resulting set $\{(x_i,y_i,\sigma_i)\}_{i=1}^n$  of
$n = N_L\times N_p$ scaled data is then ordered such that $x_i<x_{i+1}$. 
The best estimates $p_*$, $\omega_*$, and $\zeta_*$  for the critical point $p_c$ and  the scaling exponents $\bar\omega$ and $\bar \zeta$, respectively,  are obtained as the arguments which minimize the cost (or quality) function
\noindent
\begin{small}   
\begin{align}
Q(\bar{p},\bar{\omega},\bar{\zeta})=\sum^{n-1}_{i=2}\frac{w(x_i,y_i,\sigma_i|x_{i-1},y_{i-1},\sigma_{i-1},x_{i+1},y_{i+1},\sigma_{i+1})}{n-2},
\end{align}
\end{small}
where 
\begin{equation}
 w(x,y,\sigma|x',y',\sigma',x'',y'',\sigma'')=\frac{(y-\bar{y})^2}{\Delta^2}
\end{equation}
are the square distances of the point $(x,y)$ from the point $(x,\bar{y})$ obtained from a linear interpolation of the preceding and following data points, i.e., $(x',y')$ and $(x'',y'')$, respectively, in units of the error 
$\Delta$
on $y-\bar{y}$.
In formulas 
\begin{equation}
    \bar{y}=\frac{(x''-x)y'-(x'-x)y''}{x''-x'},
\end{equation}
while $\Delta$ is obtained by standard propagation of errors, i.e.,
\begin{equation}
 \Delta^2 =\sigma^2+\biggr(\frac{x''-x}{x''-x'}\sigma'\biggr)^2+ \biggr(\frac{x'-x}{x''-x'}\sigma''\biggr)^2.
\end{equation}
Error bars are assigned according to the method of Ref.~\cite{KhassehPrb}. If we have to perform the minimization of $Q$ on a set of $N_L$ curves,  we consider all $N_L$ subsets composed by $N_L-1$ different curves and estimate, for each subset, the exponents and $p_*$ as explained above. The critical exponents and $p_*$ for the whole set of data
are then estimated as the mean of the $N_L$ previous estimates. As the values extracted from each subset are expected to be correlated and an accurate determination of the resulting error would thus require a more complex analysis beyond what is needed here, we simply assign as an error on the final estimates their standard deviations instead of those of their means.

\section{Renyi-PCA entropies}
\label{app:Renyi-PCA entropies}

\begin{figure}
{\includegraphics[width=.45\textwidth]{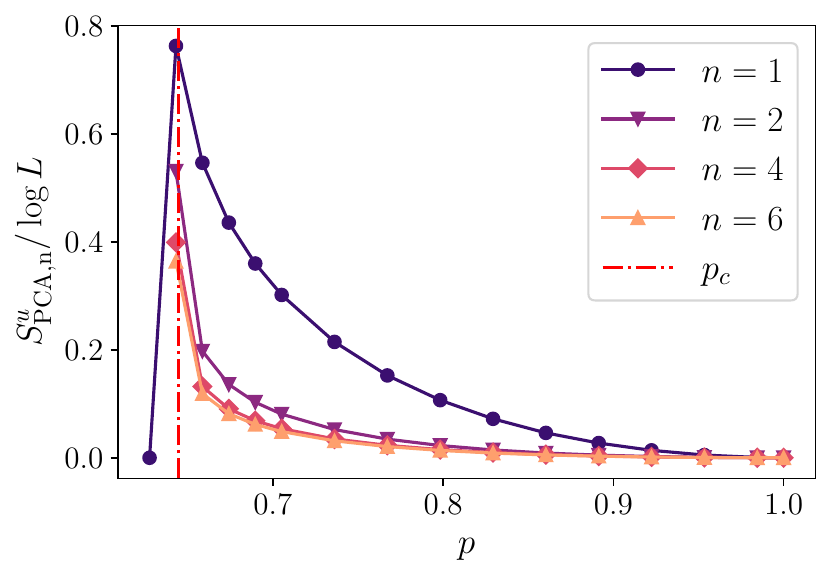}}
\caption{Uncentered Renyi-PCA entropies $S^u_{{\rm PCA},n}$ as a function of $p$ for DBP and various values of $n$. The numerical data are obtained for $L=512$, $N_r=3000$. Each data point is averaged over 5 repetitions.}
\label{fig:DP_All_Renyi_u}
\end{figure}

\begin{figure}
{\includegraphics[width=.45\textwidth]{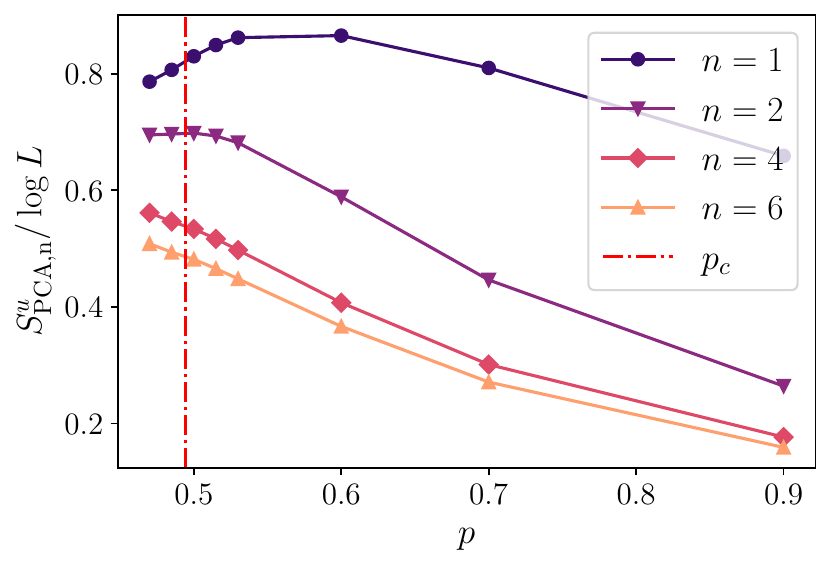}}
\caption{Uncentered Renyi-PCA entropies $S^u_{{\rm PCA},n}$  as a function of $p$ for BARWe and various values of $n$. The numerical data are obtained for $L=512$, $N_r=3000$. Each data point is averaged over 10 repetitions. }
\label{fig:PC_Entropies_Steady}
\end{figure}

In this appendix, we report the behavior of uncentered Renyi-PCA entropies $S^u_{\rm{PCA},n}$ defined in Eq.~\eqref{eq:def-Reny}. 
As discussed in Sec.~\ref{ss:entr-DP} of the main text for DBP, the peak in the PCA entropy (see Eq.~\eqref{eq:PCA-ent-def}) provides a good estimate of the critical point.
A similar behavior is displayed by $S_{\rm{PCA},n}$ for generic $n$, as shown in Fig.~\ref{fig:DP_All_Renyi_u}.
Upon increasing $n$, the values of these Renyi entropies are increasingly determined by the first few largest components of the PCA spectrum $\{\pi_i^u\}$. This indicates that, consistently with what was observed in Sec.~\ref{sec:PCA_DP_largest} most of the information about the transition is actually encoded in the behavior of the largest component.
%%%%%%%%%%%%%%%%%%%%%%%%%%%%%%%%%%%%%%%
%%%%%%%%%%%%%%%%%%%%%%%%%%%%%%%%%%%%%%%
%%
%%
As discussed in the main text in Sec.~\ref{sec:PC_largest} the case of BARWe is complicated by a richer structure in the PCA spectrum which affects also the behavior of the corresponding $S^u_{\rm {PCA},n}$.  This is shown for the various $S^u_{\rm{PCA},n}$ in
Fig.~\ref{fig:PC_Entropies_Steady}. 
Upon increasing $n$ we observe that the peak of $S_{\rm{PCA},n}$ in the active phase disappears and the curves become monotonic. While $S_{\rm PCA}$ features a local maximum above $p_c$, due to the relevant contributions coming from the various components of the spectrum beyond $\pi_1^u$, $S^u_{{\rm PCA},n}$ increasingly mirrors the behavior of $\pi_1^u$. However, even the behavior of $\pi_1^u$ does not provide as much clear information about the occurrence of the phase transition as in the case of DBP.
%%
%%
%%%%%%%%%%%%%%%%%%%%%%%%%%%%%%%%%%%%
%%%%%%%%%%%%%%%%%%%%%%%%%%%%%%%%%%%%
%%%%%%%%%%%%%%%%%%%%%%%%%%%%%%%%%%%
%%%%%%%%%%%%%%%%%%%%%%%%%%%%%%%%%%%
%%
%%

\section{Centered PCA analysis}
\label{app:Centered}

\begin{figure}
{\includegraphics[width=.45\textwidth]{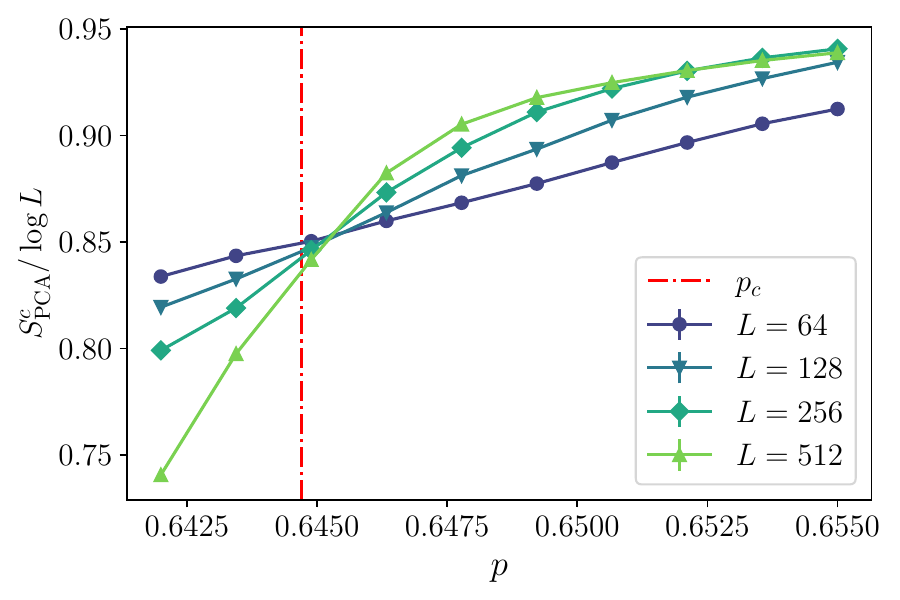}}
\caption{Centered normalized PCA entropy $S^c_{\rm PCA}$ as a function of $p$ in the vicinity of the critical point for DBP and various values of the system size $L$. The curves corresponding to the various $L$ cross each other close to criticality. Simulations correspond to $N_r=2000$ and each point is averaged over $5$ repetitions. }
\label{fig:DP_PCA_C_Zoom}
\end{figure}

\begin{figure}
{\includegraphics[width=.45\textwidth]{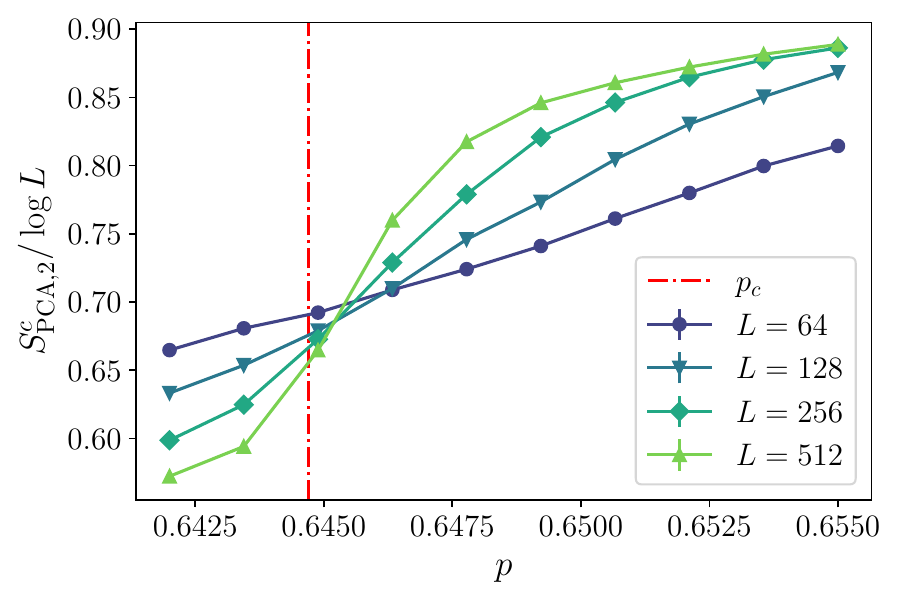}}
\caption{Centered normalized PCA-Renyi 2 entropy $S^c_{{\rm PCA},2}$ in the vicinity of the critical point for DBP and various values of the system size $L$. The curves corresponding to the various $L$ cross each other close to criticality, as it occurs for $S^c_{\rm PCA}$  in Fig.~\ref{fig:DP_PCA_C_Zoom}.Simulations correspond to $N_r=2000$ and each point is averaged over $5$ repetitions.}
\label{fig:DP_RENYI2_C_Zoom}
\end{figure}

In this appendix we report the analysis of the centered entropies, focusing on the PCA entropy $S^c_{\rm PCA}$ and on the Renyi-PCA entropy of order 2, i.e., $S^c_{{\rm PCA},2}$ defined on the basis of the centered spectrum $\{\pi_i^c\}$ in Eqs.~\eqref{eq:PCA-ent-def} and \eqref{eq:def-Reny}, respectively. 
We first discuss these quantities for the case of DBP, finding that they are also 
informative about the transition. In particular, the centered PCA entropy $S^c_{\rm PCA}$ can be used to locate the 
critical point. This is shown in Fig.~\ref{fig:DP_PCA_C_Zoom}, 
which reports $S^c_{\rm{PCA}}$ as a function of $p$ close to the expected critical value, and for various values of the system size $L$. Differently from what happens for $S_{\rm{PCA}}^u$, the centered PCA entropy $S^c_{\rm{PCA}}$ does not display a peak close to the critical point, but only
features a crossing of the various curves for different values of system size. Similar behavior is observed also in the PCA-Renyi2 entropy $S^c_{{\rm PCA},2}$, as shown in Fig.~\ref{fig:DP_RENYI2_C_Zoom}. 
%%
%%
%%%%%%%%%%%%%%%%%%%%%%%%%%%%%%%%%%%%%%%%%%%
%%%%%%%%%%%%%%%%%%%%%%%%%%%%%%%%%%%%%%%%%%%
%%%%%%%%%%%%%%%%%%%%%%%%%%%%%%%%%%%%%%%%%%%
%%%%%%%%%%%%%%%%%%%%%%%%%%%%%%%%%%%%%%%%%%%
%%
%%
%%%%%%%%%%%%%%%%%%%%%%%%%%%%%%%%%%%%%%%%%%%
%%%%%%%%%%%%%%%%%%%%%%%%%%%%%%%%%%%%%%%%%%%
%%%%%%%%%%%%%%%%%%%%%%%%%%%%%%%%%%%%%%%%%%%
%%%%%%%%%%%%%%%%%%%%%%%%%%%%%%%%%%%%%%%%%%%
%%
%%

\begin{figure}
{\includegraphics[width=.45\textwidth]{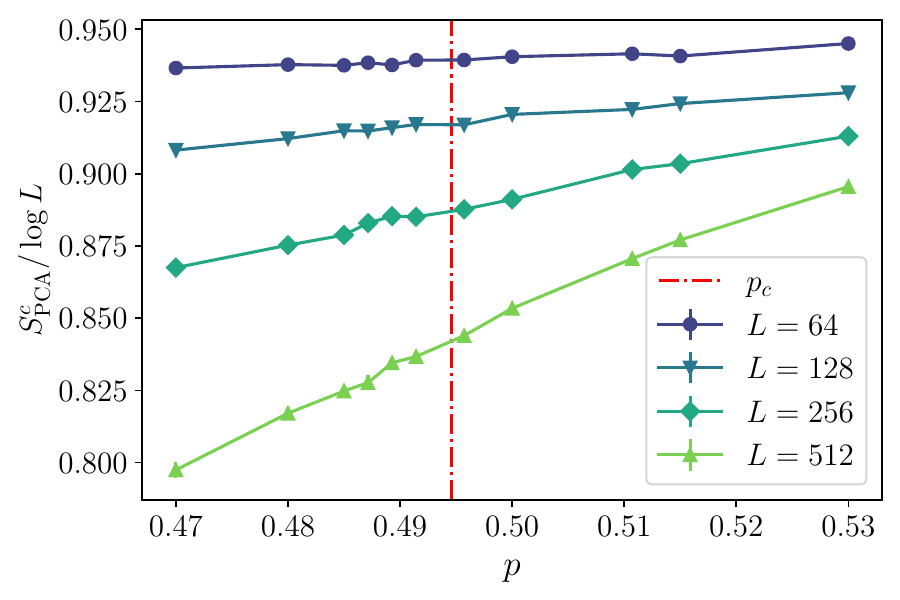}}
\caption{Centered normalized PCA entropy $S^c_{\rm PCA}$ as a function of $p$ in the vicinity of the critical point for BARWe and various values of the system size $L$. Differently from the case of directed percolation shown in Fig.~\ref{fig:DP_PCA_C_Zoom}, no crossing is observed.Simulations correspond to $N_r=3000$, and points are averaged over 10 repetitions.}
\label{fig:PC_PCA_C_Zoom}
\end{figure}

\begin{figure}
{\includegraphics[width=.45\textwidth]{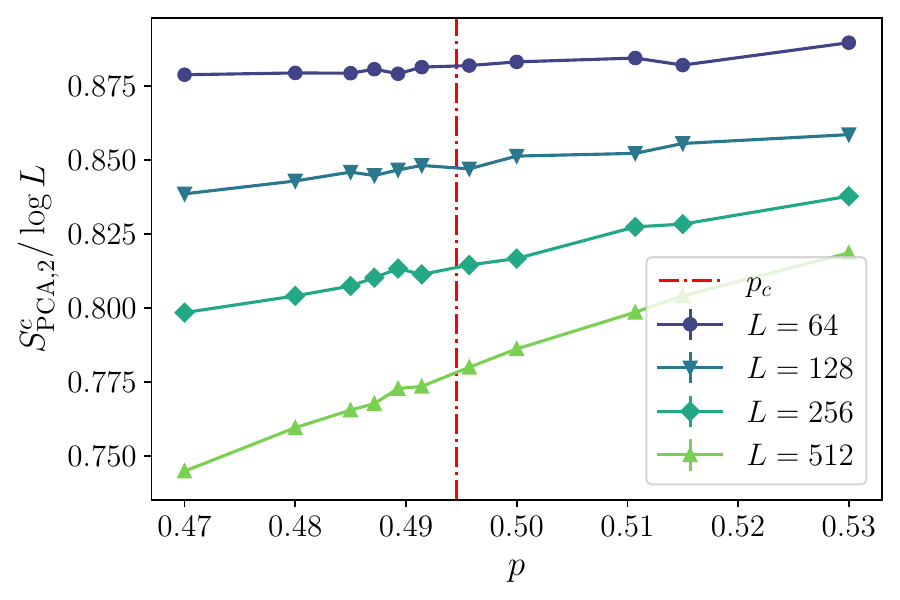}}
\caption{Centered normalized PCA-Renyi 2 entropy $S^c_{{\rm PCA},2}$ in the vicinity of the critical point for BARWe and various values of the system size $L$. Differently from the case of directed percolation shown in Fig.~\ref{fig:DP_RENYI2_C_Zoom}, no crossing is observed. Simulations correspond to $N_r=3000$, and points are averaged over 10 repetitions.}
\label{fig:PC_RENYI2_C_Zoom}
\end{figure}

The behavior of $S^c_{{\rm PCA}}$ and $S^c_{{\rm PCA},2}$ for the BARWe is shown in Figs.~\ref{fig:PC_PCA_C_Zoom} and \ref{fig:PC_RENYI2_C_Zoom}, respectively. In this case, no crossing is observed in these quantities, when normalized in the same fashion, i.e., by $\log L$. 
This might be related to the richer structure that the PCA spectrum has for BARWe, which was also observed for the corresponding uncentered quantites discussed in Sec.~\ref{sec:PC_largest} of the main text.

%%
%%%%%%%%%%%%%%%%%%%%%%%%%%%%%%%%%%%%%%%%%%%
%%%%%%%%%%%%%%%%%%%%%%%%%%%%%%%%%%%%%%%%%%%
%%%%%%%%%%%%%%%%%%%%%%%%%%%%%%%%%%%%%%%%%%%
%%%%%%%%%%%%%%%%%%%%%%%%%%%%%%%%%%%%%%%%%%%
%%
%%
%%%%%%%%%%%%%%%%%%%%%%%%%%%%%%%%%%%%%%%%%%%
%%%%%%%%%%%%%%%%%%%%%%%%%%%%%%%%%%%%%%%%%%%
%%%%%%%%%%%%%%%%%%%%%%%%%%%%%%%%%%%%%%%%%%%
%%%%%%%%%%%%%%%%%%%%%%%%%%%%%%%%%%%%%%%%%%%
%%

\begin{figure}[t]
{\includegraphics[width=.45\textwidth]{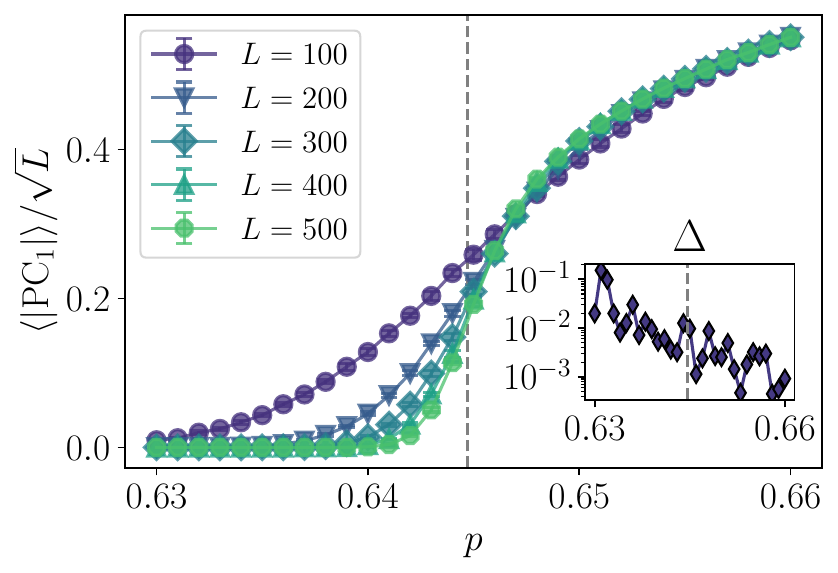}}
\caption{First QPC $\langle |\mathrm{PC}_1|\rangle$ for DBP in the stationary state. We take $N_r=1000$ for each value of $p$. The inset shows the relative difference $\Delta$ defined in Eq.~\eqref{eq:app-def-delta}, between the density $\rho$ of active sites and the first QPC for 
$L=100$ (similar behavior is observed for other values of $L$).
This demonstrates a quantitative agreement between the two quantities across the considered range of values of the control parameter. The dashed line indicates the position of the critical point $p_c\approx 0.6447$.}  
\label{fig:qpc1_static}
\end{figure}

\section{Quantified principal components}
\label{app:Complementary}

The analyses presented in the main text of this work and in the previous appendices where based on doing a PCA on the data matrix $\mathbf{X}$ (see the definition in Eq.~\eqref{eq:def-X}), which collects various configurations of the system at a given value of $p$ (in the steady-state) or at a given $p$ and time $t$ (during relaxation). Here, we discuss a complementary type of analysis in which one collects all available data into a single data matrix. Let us illustrate the procedure for the case of datasets in the steady state. 
For a fixed system size $L$, we collect the data matrices $\mathbf{X}$ (constructed as explained in Sec.~\ref{sec:methods}) corresponding to each single value of $p$ which, for clarity, we denote by $\mathbf{X}_p$. Then we consider the dataset formed by the matrices $\mathbf{X}_p$ corresponding to all the values of $p$ which are considered in the analysis, i.e., $\{\mathbf{X}_p\}$ and we collect all these observations into a single data matrix, which we denote by $\mathbf{Y}$. 
This data matrix has the same number of columns (corresponding to the number of lattice sites) as the individual matrices $\mathbf{X}_p$. However, the number of rows is now $N_r \times N_p$ (instead of $N_r$), where $N_p$ is the number of values of the control parameter $p$ probed by the simulations. 
We then performed a column-centered PCA on $\mathbf{Y}$ (as in Eq.~\eqref{eq:centering} with $\mathbf{X} \mapsto \mathbf{Y}$ and $N_r \mapsto N_r \times N_p$) and looked at the information contained in the eigenvectors of the covariance matrix of $\mathbf{Y}$ (see Eq.~\eqref{eq:def-cov}). 
More concretely,  following Ref.~\cite{PhysRevE.95.062122}, we define the \emph{quantified principal components} (QPCs) as the average of the inner product of the data points corresponding to a given $p$ with the $L$-dimensional normalized eigenvectors $\vec{w}_i$ of the covariance matrix associated to $\mathbf{Y}$. Accordingly, the QPCs are given by
    \begin{equation}
    \label{eq:qpc}
        \langle |\mathrm{PC}_i| \rangle(p) := \frac{1}{N_r} \sum_{\mu=1}^{N_r} |\vec{x}^{(\mu)}(p) \cdot \vec{w}_i|,
    \end{equation}
where $\vec{x}^{(\mu)}(p)$ are the $L$-dimensional row vectors in $\mathbf{X}_p$.

\begin{figure}[t]
{\includegraphics[width=.45\textwidth]{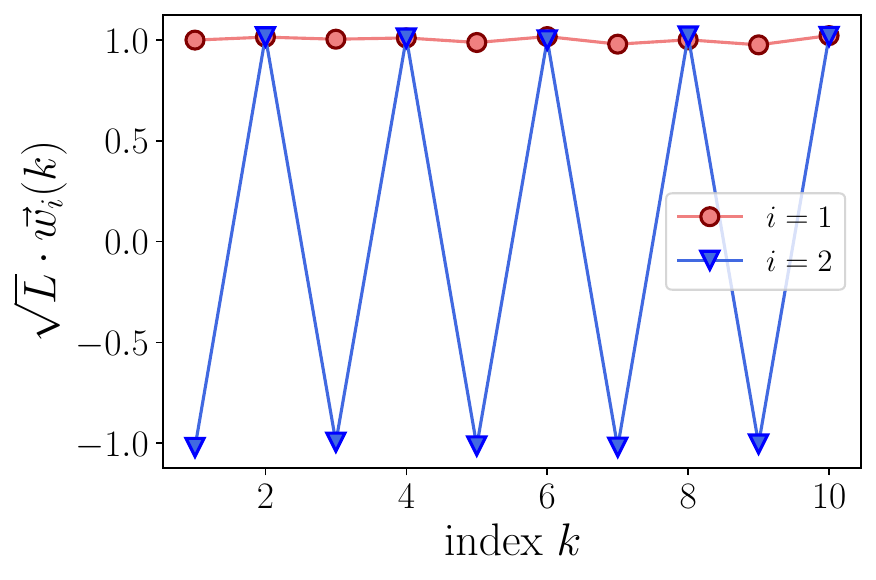}}
\caption{Entries of the first two normalized eigenvectors of $\mathbf{Y}$, as a function of the entry index $k$, for the steady state analysis of DBP with $L=100$. Only the first 10 entries of these $L$-dimensional vectors are shown here for clarity, however, the same structure is observed for the remaining entries of both eigenvectors. The same behavior occurs irrespective of the value of $L$ (up to numerical variations).}  
\label{fig:wcomponents}
\end{figure}
%%%%%%%%%%%%%%%%%%%%%%%%%%%%%%%%%%%%%%%%%%%
%%%%%%%%%%%%%%%%%%%%%%%%%%%%%%%%%%%%%%%%%%%

\begin{figure}[t]
{\includegraphics[width=.47\textwidth]{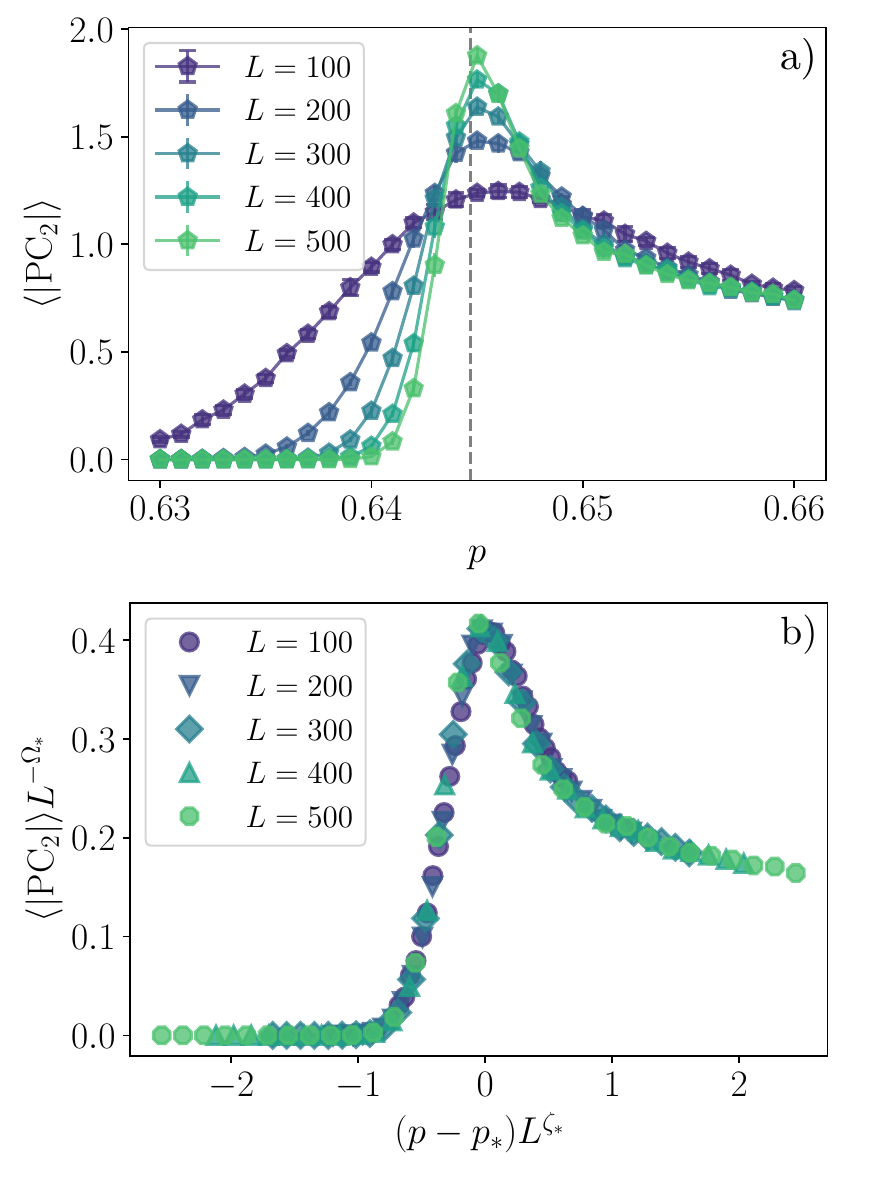}}
\caption{(a) Second QPC $\langle |\mathrm{PC}_2|\rangle$ for DBP in the stationary state, which shows a clear signature of the transition point as the system size is increased. The various remaining parameters take the same values as in Fig.~\ref{fig:qpc1_static}. (b) Finite-size scaling of the second QPC in the stationary state. The obtained critical point and exponents that realize the data collapse are given by $p_*=0.64484(5)$, $\Omega_*=0.261(4), \zeta_*=0.90(1)$. }  
\label{fig:qpc2_static}
\end{figure}

The first two QPCs, i.e., $ \langle |\mathrm{PC}_1| \rangle$ and  $\langle |\mathrm{PC}_2| \rangle$, for the stationary behavior (with the system configurations being sampled at a fixed time $t=L^z$) of DBP, are shown in Figs.~\ref{fig:qpc1_static} and \ref{fig:qpc2_static}, respectively (the corresponding results are averaged over 10 independent repetitions). 
For each value of $p$, we used $N_r=1000$. 
These figures reveal that the critical properties of the model are contained  also in the eigenvectors of the covariance matrix associated to the data matrix $\mathbf{Y}$. 
In particular, we observe that the first QPC, rescaled by a factor of $\sqrt{L}$, captures quantitatively the behavior of the order parameter $\rho$  (see Fig.~\ref{fig:qpc1_static}). This can be understood from the numerical observation that, for the system under consideration, the structure of the first eigenvector $\vec{w}_1$ is roughly homogeneous, namely,  $\vec{w}_1\approx \frac{1}{\sqrt{L}}(1,1, \dots, 1)$. This is shown in Fig.~\ref{fig:wcomponents}, which reports the first 10 entries of $\vec{w}_1$ and $\vec{w}_2$. Hence, because of the definition of $\langle |\mathrm{PC}_1|\rangle$ in Eq.~\eqref{eq:qpc}, it is clear its connection to the order parameter, namely,
    \begin{equation}
    \label{eq:qpc_1_rho}
        \langle |\mathrm{PC}_1| \rangle(p) \approx \frac{1}{N_r} \frac{1}{\sqrt{L}} \sum_{\mu=1}^{N_r} \sum_{i=1}^L x^{(\mu)}_{i}(p) \approx \sqrt{L} \rho(p) ,
    \end{equation}
where $\rho(p)$ is the stationary density at a given $p$.

In the inset of Fig.~\ref{fig:qpc1_static}, we plot the relative 
difference $\Delta$ between the particle density $\rho$ and $\langle |\mathrm{PC}_1|\rangle/\sqrt{L}$, i.e.,  
\begin{equation}
    \Delta := \left|\frac{\rho - \langle |\mathrm{PC}_1|\rangle/\sqrt{L}}{\rho}\right|,
    \label{eq:app-def-delta}
\end{equation} for $L=100$ (similar behavior is observed for other values of $L$). This difference turns out to be always rather small, independently of the value of $p$, as expected from Eq.~\eqref{eq:qpc_1_rho}.

The second QPC $\langle |\mathrm{PC}_2|\rangle$ in Fig.~\ref{fig:qpc2_static}(a) clearly signals the onset of the phase transition in a fashion that closely resembles the behavior of a susceptibility for a system at equilibrium featuring a continuous (second-order) phase transition. Note that $\langle |\mathrm{PC}_2|\rangle$ %Such, quantity, whose definition 
reminds of a staggered density (see $\vec{w}_2$ in Fig.~\ref{fig:wcomponents}) and, to our knowledge, it has not been previously examined for the DBP. Accordingly, the fact discussed below that it displays an interesting scaling behavior can be considered a genuine discovery of the PCA.
In order to test if this quantity has good scaling properties we perform a finite-size scaling with the ansatz
\begin{equation}
    \langle |\mathrm{PC}_2| \rangle(p) = L^{\Omega} h((p-p_c)L^\zeta).
\end{equation}
As shown in Fig~\ref{fig:qpc2_static}(b) $\langle |\mathrm{PC}_2| \rangle$ exhibits good scaling properties, which is made explicit by the collapse of data for different system size onto a master curve $h$. The values of the critical point and of the exponents that yield the best data collapse are $p_*=64484(5)$, $\Omega_*=0.261(4), \zeta_*=0.90(1)$. Remarkably $p_*$ and $\zeta_*$ are good estimates for the expected values of $p_c$ and $1/\nu_\perp$. Concerning $\Omega_*$, it appears that $\Omega_* \approx -\omega_*$ (with $\omega_*$ reported in the main text in Sec. \ref{sec:PCA_DP}).

%%
%%
%%
%%%%%%%%%%%%%%%%%%%%%%%%%%%%%%%%%%%%%%%%%%%
%%%%%%%%%%%%%%%%%%%%%%%%%%%%%%%%%%%%%%%%%%%

%%%%%%%%%%%%%%%%%%%%%%%%%%%%%%%%%%%%%%%%%%%
%%%%%%%%%%%%%%%%%%%%%%%%%%%%%%%%%%%%%%%%%%%
%%
%%

\begin{figure}[h]
{\includegraphics[width=.49\textwidth]{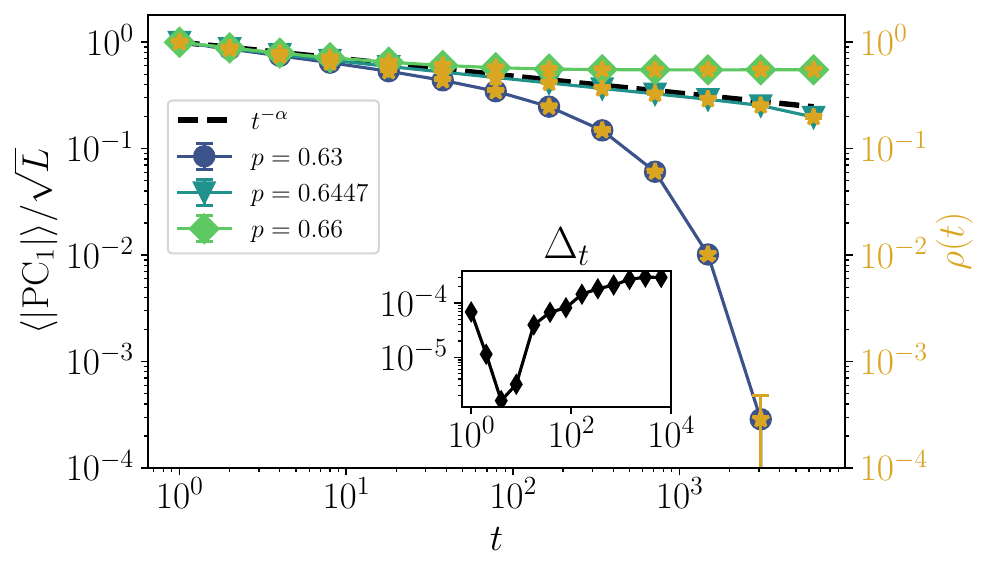}}
\caption{First QPC $\langle |\mathrm{PC}_1|\rangle$ for DBP during relaxation from a fully active initial state, for  $L=256$, and $N_r=1000$ (for each value of $t$). 
The $\star$ markers show the density $\rho(t)$ of active sites (the corresponding scale is shown along the vertical axis on the right side of the plot), showing a quantitative agreement between the two quantities. The inset shows the value $\Delta_t$  of the relative difference $\Delta$ (see Eq.~\eqref{eq:app-def-delta}) between the density $\rho(t)$ of active sites and $\langle |\mathrm{PC}_1|\rangle/\sqrt{L}$ at $p=0.6447$.
This demonstrates a quantitative agreement between the two quantities across the considered range of evolution times.}  
\label{fig:qpc1_dynamical}
\end{figure}

The same analysis can also be applied to the data concerning the time evolution of the system, where now we collect all the available data at various times during the evolution (at fixed $p$). 
Figure~\ref{fig:qpc1_dynamical} shows $\langle |\mathrm{PC}_1|\rangle$ as a function of $t$, for various values of $p$ and $L=256$. At each time $t$, we collect $N_r=1000$ configurations. We show the results after averaging over 10 independent runs. This plot also shows the corresponding value of the time-dependent density of active sites $\rho(t)$ ($\star$ markers), which is practically indistinguishable from $\langle |\mathrm{PC}_1|\rangle$ (rescaled by $\sqrt{L}$). This also follows from the structure of the first normalized eigenvector $\vec{w}_1$, which just as for the static case is roughly homogeneous, i.e., $\vec{w}_1\approx \frac{1}{\sqrt{L}}(1,1, \dots, 1)$. 
The inset in this figure shows the time-dependent value $\Delta_t$ of the quantity $\Delta$ defined in Eq.~\eqref{eq:app-def-delta}, plotted as a function of time $t$, for the data corresponding to $p=0.6447$. This relative difference turns out to be smaller than $\approx 10^{-3}$.
It is thus clear that even when considered in the time domain, $\langle |\mathrm{PC}_1|\rangle$ reproduces quantitatively the behavior of the order parameter. 
%%%%%%%%%%%%%%%%%%%%%%%%%%%%%%%%%%%%%%%%%%%
%%%%%%%%%%%%%%%%%%%%%%%%%%%%%%%%%%%%%%%%%%%
%%%%%%%%%%%%%%%%%%%%%%%%%%%%%%%%%%%%%%%%%%%
%%%%%%%%%%%%%%%%%%%%%%%%%%%%%%%%%%%%%%%%%%%
%%
%%
\end{document}